\documentclass[prx,aps,twocolumn]{revtex4-1}

\usepackage[pdftex]{graphicx}
\usepackage{amsbsy,amssymb,amsmath,bm,mathtools}

\usepackage{xspace}
\usepackage{bm}
\usepackage{color}

\usepackage{url}
\usepackage[section]{placeins}
\usepackage[colorlinks=true,linkcolor=blue,citecolor=blue]{hyperref}

\begin{document}


\title{Descriptors for Machine Learning Model of Generalized Force Field \\ in Condensed Matter Systems}

\author{Puhan Zhang}
\affiliation{Department of Physics, University of Virginia, Charlottesville, VA 22904, USA}

\author{Sheng Zhang}
\affiliation{Department of Physics, University of Virginia, Charlottesville, VA 22904, USA}

\author{Gia-Wei Chern}
\affiliation{Department of Physics, University of Virginia, Charlottesville, VA 22904, USA}

\date{\today}

\begin{abstract}
We outline the general framework of machine learning (ML) methods for multi-scale dynamical modeling of condensed matter systems, and in particular of strongly correlated electron models. Complex spatial temporal behaviors in these systems often arise from the interplay between quasi-particles and the emergent dynamical classical degrees of freedom, such as local lattice distortions, spins, and order-parameters. Central to the proposed framework is the ML energy model that, by successfully emulating the time-consuming electronic structure calculation, can accurately predict a local energy based on the classical field in the intermediate neighborhood.  In order to properly include the symmetry of the electron Hamiltonian, a crucial component of the ML energy model is the descriptor that transforms the neighborhood configuration into invariant feature variables, which are input to the learning model. A general theory of the descriptor for the classical fields is formulated, and two types of models are distinguished depending on the presence or absence of an internal symmetry for the classical field. Several specific approaches to the descriptor of the classical fields are presented. Our focus is on the group-theoretical method that offers a systematic and rigorous approach to compute invariants based on the bispectrum coefficients. We propose an efficient implementation of the bispectrum method based on the concept of reference irreducible representations.  Finally, the implementations of the various descriptors are demonstrated on well-known electronic lattice models. 
\end{abstract}

\maketitle

\section{Introduction}
\label{sec:intro}

Machine learning (ML) is emerging as a new paradigm for scientific research and engineering~\cite{dunjko18,kalinin15,carleo19,radovic18,sarma19,baron19,morgan20,butler18,meuwly21,keith21,greener21,bedolla21,libbrecht15,ourmazd20}. In particular, ML methods are increasingly employed in recent years to drastically speedup various computational tasks in quantum chemistry and materials science~\cite{rupp12,snyder12,brockherde17,schutt19,wang19,tsubaki20,burkle21,huang17,liu17}.  A ML model can be viewed as a complex high-dimensional function with numerous tunable parameters. Highly efficient methods have been developed to optimize these model parameters from large number of training dataset. Among the various ML models, deep neural networks (NN)~\cite{schmidhuber14,lecun15} represent the most powerful and versatile tools, which, in principle, can approximate any continuous function with arbitrary accuracy~\cite{cybenko89,hornik89,barron93}.  One of the most remarkable applications along this line is the development of ML models that can emulate the time-consuming first-principles electronic structure calculations based on, e.g. the density functional theory (DFT), thus significantly surpassing the size and time scales accessible to such accurate methods.  
Notably, the advent of interatomic potentials based on ML models has made it possible to perform large-scale molecular dynamics (MD) simulations with the accuracy of DFT and beyond~\cite{behler07,bartok10,li15,botu17,li17,smith17,zhang18dp,behler16,deringer19,mcgibbon17,suwa19,mueller20,noe20}.

The success of ML methods in quantum chemistry and quantum MD simulations has motivated similar applications in condensed-matter physics. For example, the utilization of ML model to emulate the complicated many-body calculation could potentially offer the tantalizing potential for accurate multi-scale dynamical modeling of interacting electron systems. A particularly important application is the large-scale simulations of the spatio-temporal dynamics of complex patterns that are prevalent in correlated electron materials~\cite{dagotto_book,dagotto05,moreo99,mathur03,kivelson98,tranquada95,hanaguri04,vershinin04,pan01,lang02,chen11}. The intriguing nanoscale textures in such systems are believed to arise from the nontrivial interplay between quasi-equilibrium electrons and emergent classical fields or bosonic degrees of freedom whose dynamics is much slower than the relaxation of electrons. An example of such slow dynamical variables is the magnetic moments associated with localized $d$ or $f$ electrons immersed in the Fermi sea of fast-moving conducting electrons in the s-d or double-exchange model~\cite{zhang04,yunoki98,zener51,anderson55,degennes60}. Another representative case is the order-parameter fields which couple to equilibrium quasi-particles in a symmetry-breaking phase~\cite{fradkin_book,gruner88,gruner94,fradkin15,chern18}. The well separated time scales in such electron models is similar to the Born-Oppenheimer approximation underlying the {\em ab-initio} molecular dynamics methods~\cite{marx09,iftimie05}. Instead of the atomic dynamics, the goal then is to model the adiabatic time evolution of slow dynamical variables such as local spins and order-parameter fields under the influence of the fast electron degrees of freedom.

Conventionally, an empirical or effective energy model as a function of the classical fields, such as an effective classical spin Hamiltonian or Ginzburg-Landau energy functional, is employed for large-scale dynamics simulations~\cite{onuki02,bray94}. Special care is taken to properly incorporate the symmetry of the original quantum Hamiltonian into the effective model. However, while the classical energy models coupled with phenomenological dynamics capture some universal features qualitatively, such empirical approach lacks the predictive power. Moreover, the effective classical energy, which can be viewed as integrating out the electrons beforehand, fails to describe the subtle interplay between the electron and the classical degrees of freedom during the dynamical evolution.

In order to more accurately simulate the dynamics of the classical fields, one needs to integrate out the fast electrons or quasi-particles on the fly. This means that the fermionic Hamiltonian, characterized by the instantaneous classical fields, needs to be solved at every time-step of the dynamical simulations. Compared with the classical energy model discussed above, this {\em quantum} dynamical approach is similar in spirit to the quantum MD methods in which the atomic forces are obtained by solving, e.g. the self-consistent Kohn-Sham equation at every time-step~\cite{marx09,iftimie05}. Naturally, the huge computational overhead due to the repeated electronic structure calculations significantly limits the system size and simulation time accessible by such quantum approaches.  For electron systems with e.g. Hubbard-type interactions, more sophisticated, hence more time-consuming, many-body methods such as the dynamical mean-field theory~\cite{georges96,kotliar06}, density-matrix renormalization group~\cite{white92,schollwoeck04}, or quantum Monte Carlo~\cite{ceperley86}, are required to properly include the strong electron correlation effects.

As mentioned above, the modern ML methods offer a promising solution to this computational difficulty in multi-scale quantum dynamical modeling of classical fields, as demonstrated by the ML-based interatomic potential for quantum MD simulations. Indeed, recent works~\cite{zhang20,zhang21} have demonstrated the use of deep-learning NN models to enable large-scale quantum Landau-Lifshitz-Gilbert dynamics simulation of phase separation phenomena in a correlated electron system known as the double-exchange model~\cite{yunoki98,zener51,anderson55,degennes60}. Moreover, ML energy model was used to achieve large-scale quantum kinetic Monte Carlo simulations which reveals unexpected phase ordering dynamics in the Falicov-Kimball model~\cite{zhang21a}, another canonical example of correlated electron systems~\cite{falicov69,freericks03}. The classical degrees of freedom in the former case are local magnetic moments, while they are equivalent to a classical lattice gas mode used to describe the heavy $f$ electrons in the latter case. In both applications, the electronic subsystem is described by quadratic fermionic Hamiltonians, which can be exactly diagonalized. These pilot studies, however, demonstrate the plausibility of applying ML methods to lattice models with strong electron-electron interactions, such as the Hubbard-type models.  The general framework of the ML energy model for multi-scale dynamical simulations is discussed in Sec.~\ref{sec:framework}.

It is worth noting that the ML approach to the multi-scale modeling discussed above is essentially to develop a classical energy model based on the superb approximation power of modern learning models such as the NN. Large-scale dynamical simulations are possible mainly because of the efficiency of computing forces based on the classical effective energy. However, even with the general approximation capability of ML methods, it is not guaranteed that the symmetry of the original electron Hamiltonian can be properly included in the effective ML model. In order to ensure the symmetry properties of energy model, one needs to first construct a proper representation of the classical field configuration to be used as input to the learning models. A good representation is {\em invariant} with respect to transformations of the symmetry group of the classical fields as well as those of the lattice point group. This crucial step of the ML model, namely the construction of the proper representation, is often referred to as feature engineering and the resultant feature variables, also called the generalized coordinates, are termed a descriptor~\cite{ghiringhelli15,bartok13,himanen20}.

Similar issues have also arisen in the context of ML interatomic potentials for quantum MD simulations. There, a proper descriptor of the atomic configuration should be invariant under rotational and permutational symmetries, while retaining the faithfulness of the Cartesian representation. Over the past decade, a number of descriptors have been proposed together with the learning models based on them~\cite{rupp12,behler07,behler11,bartok10,bartok13,zhang18dp,ghiringhelli15,bartok13,himanen20,steinhardt83,behler11,shapeev16,drautz19,hansen15,faber15,huo18,ma19}. The bond-order parameters, originally developed to characterize short-range structural order in liquid and glasses~\cite{steinhardt83}, are one example. A relatively simple approach is to use the ordered eigenvalues of the correlation matrix, such as the Coulomb or Edward sum matrices, as the descriptor~\cite{rupp12}. Another example, which is physically intuitive, is the atom-centered symmetry functions (ACSFs) built from the two-body (relative distances) and three-body (relative angles) invariants of the atomic configurations. Because of its simplicity and flexibility, the ACSF descriptor is widely used in various learning models~\cite{behler07,behler11}. The group-theoretical method, on the other hand, offers a more controlled approach to the construction of atomic representation based on the power-spectrum and bispectrum coefficients~\cite{bartok10,bartok13}. In addition to these relatively well-established methods, other notable descriptors include moment tensor potential~\cite{shapeev16}, atomic cluster expansion~\cite{drautz19}, and deep-potential representation~\cite{zhang18dp}. It is worth noting that the research of atomic descriptor is an active ongoing field.

In this paper, we develop a general theory of the descriptors for the classical fields in condensed matter systems, with a special focus on the lattice models. Several specific approaches are also presented; some are motivated by and generalized from the atomic descriptors discussed above. Our main focus, however, is on the group-theoretical method, which can in principle provide a faithful representation of the local classical fields. The resultant descriptor in terms of the bispectrum coefficients of the irreducible representations (IRs) of the lattice point group is rigorous, but over-complete and cumbersome to implement. We next discuss  the concept of the reference IRs which can significantly simplify the implementation of the bispectrum descriptor. The proposed descriptors are then applied to lattice models with varying symmetries and complexity of the classical fields.

The rest of the paper is organized as follows. In Sec.~\ref{sec:framework}, we outline the framework of utilizing ML energy model to achieve multi-scale dynamical modeling of condensed matter systems. We also discuss the similarities and differences between descriptor for ML-based quantum MD methods and that of classical degrees of freedom in lattice electronic models. Sec.~\ref{sec:descriptor} presents a general formulation of the descriptors for the dynamical classical fields.  While the majority of the discussion is on the group-theoretical method for the bispectrum coefficients, we also present other physically intuitive descriptors motivated by the studies of ML-based interatomic potentials for MD simulations. In Sec.~\ref{sec:scalar}, we demonstrate the bispectrum descriptor, together with the idea of reference IRs, to lattice models with a simple scalar classical field representing local breathing-type lattice distortions. Sec.~\ref{sec:JT} discusses the case of cooperative Jahn-Teller coupling which is an example of doublet classical field that transforms simultaneously with the point group.  Sec.~\ref{sec:spin} is devoted to the systems with a classical vector fields such as the local magnetic moments in double-exchange models. We demonstrate how to properly account for the global rotation symmetry on top of the discrete lattice symmetry. We conclude our work in Sec.~\ref{sec:conclusion}.

\section{General framework}
\label{sec:framework}

The rich and complex behaviors of several correlated electron systems arise from the emergence of slow classical degrees of freedom which couple to the electron liquid with a relatively short relaxation time. These classical variables could arise from the local lattice distortions or displacements which couples to electrons through deformation potential in, e.g. adiabatic Holstein or Jahn-Teller models. They could also correspond to local magnetic moments associated with localized core electrons in the double-exchange system~\cite{yunoki98,zener51,anderson55,degennes60}. The classical field can also represent the collective degrees of freedom such as order parameters of symmetry breaking phases~\cite{fradkin_book,gruner88,gruner94,fradkin15,chern18}, or amplitudes of slave bosons in Gutzwiller theory of Mott metal-insulator transitions~\cite{wen_book,kotliar86,lanata17}. 

In the following, we denote these emergent classical fields by an array of classical variables associated at every sites of the lattice:
\begin{eqnarray}
	\label{eq:Phi_vector}
	\bm\Phi(\mathbf r_i) = \bm\Phi_i = (\Phi_{i, 1}, \Phi_{i, 2}, \cdots, \Phi_{i, M}) 
\end{eqnarray}
We then consider the following general fermionic Hamiltonian characterized by the classical fields:
\begin{eqnarray}
	\label{eq:H1}
	\hat{\mathcal{H}} &=&  \sum_{mn}\sum_{\alpha\beta} t_{m\alpha, n\beta}(\{ \bm\Phi_i\}) \hat{c}^\dagger_{m \alpha} \hat{c}^{\,}_{n \beta} \\
	& & + \sum_{mnkl} \sum_{\alpha\beta\gamma\delta} v_{m\alpha, n\beta, k\gamma, l\delta}(\{\bm\Phi_i\}) 
	\hat{c}^\dagger_{m\alpha} \hat{c}^{\dagger}_{n\beta} \hat{c}^{\,}_{l\gamma} \hat{c}^{\,}_{k\delta}, \nonumber
\end{eqnarray}
where $\hat{c}^\dagger_{i, \alpha}$ is the creation operator of electron with quantum number $\alpha$ at site-$i$.
In the following we use the latin letters $i,j,k, \cdots$ to denote the lattice sites, and the Greek letters $\alpha,\beta, \cdots$ for internal degrees of freedom, such as spins and orbitals, of the electrons. $t$ and $v$ are electron hopping and interaction coefficients that depend on $\bm\Phi_i$. 
Finally, there is in general also a ``classical" potential energy $\mathcal{V}(\{\bm\Phi_i\})$ for the classical fields, which is independent of the electrons.  

Here we are interested in the adiabatic dynamics of such lattice fermion systems with mixed quantum (electron) and classical degrees of freedom. The adiabatic approximation, which is similar to the Born-Oppenheimer approximation in quantum molecular dynamics (MD)~\cite{marx09,iftimie05}, is based on a well separation of time scales for the electrons and the classical variables. It assumes that the relaxation of electrons is much faster than the dynamical evolution of the classical variables. As a result, the time evolution of the classical $\bm\Phi_i$ field is determined by the quasi-equilibrium electronic state of the instantaneous Hamiltonian. Specifically, this electronic state is represented by the many-body density matrix
\begin{eqnarray}
	\hat{\rho}_e\left(\{\bm\Phi_i\} \right) = \exp\bigl[-\beta \hat{\mathcal{H}}\left( \{ \bm \Phi_i \} \right) \bigr] / \mathcal{Z},
\end{eqnarray}
where $\beta = 1/k_B T$ is the inverse temperature, and $\mathcal{Z} = {\rm Tr}e^{-\beta \hat{\mathcal{H}}}$ is the instantaneous partition function of the electrons. The evolution of the classical fields is governed by dynamical equations ranging from phenomenological relaxational and Metropolis/Glauber-type dynamics to Newton/Langevin equation of motion or Landau-Lifshitz-Gilbert (LLG) equation. For example, one of the simple pure relaxation equation is the time-dependent Ginzburg-Landau (TDGL) equation~\cite{onuki02,bray94}, also known as the model-A dynamics~\cite{hohenberg77}:
\begin{eqnarray}
	\label{eq:TDGL}
	\frac{\partial \bm \Phi_i}{\partial t} = -\lambda \frac{\partial {E}}{\partial \bm \Phi_i},
\end{eqnarray}
where $\lambda$ is a dissipation constant, and the effective energy is obtained from the expectation value of the instantaneous Hamiltonian, 
\begin{eqnarray}
	\label{eq:E1}
	{E} = \langle \hat{\mathcal{H}}\bigl(\{\bm\Phi_i \} \bigr) \rangle = {\rm Tr}\bigl( \hat{\rho}_e \hat{\mathcal{H}} \bigr).
\end{eqnarray}
In general, the dynamics of the classical fields is determined by the ``forces", which is the derivative of the effective energy with respect to the classical variables. An example is given by the right-hand side of the TDGL equation~(\ref{eq:TDGL}). Given this electron density matrix, one can then compute the generalized electronic forces acting on the classical variables:
\begin{eqnarray}
	\label{eq:forces}
	\bm{\mathcal{F}}_i = - \frac{\partial \langle \hat{\mathcal{H}} \rangle}{\partial \bm\Phi_i} = - \frac{\partial {\rm Tr}\bigl( \hat{\rho}_e \hat{\mathcal{H}} \bigr)}{\partial \bm\Phi_i}.
\end{eqnarray}
It is worth noting that these forces have to be computed at every time-step of the dynamical simulations. Finally, the ``classical" potential energy $\mathcal{V}(\{\bm\Phi_i\})$ also contributes to the force $-\partial \mathcal{V} / \partial \bm\Phi_i$, which can be easily computed and included in the equation of motion.

\begin{figure*}
\includegraphics[width=1.95\columnwidth]{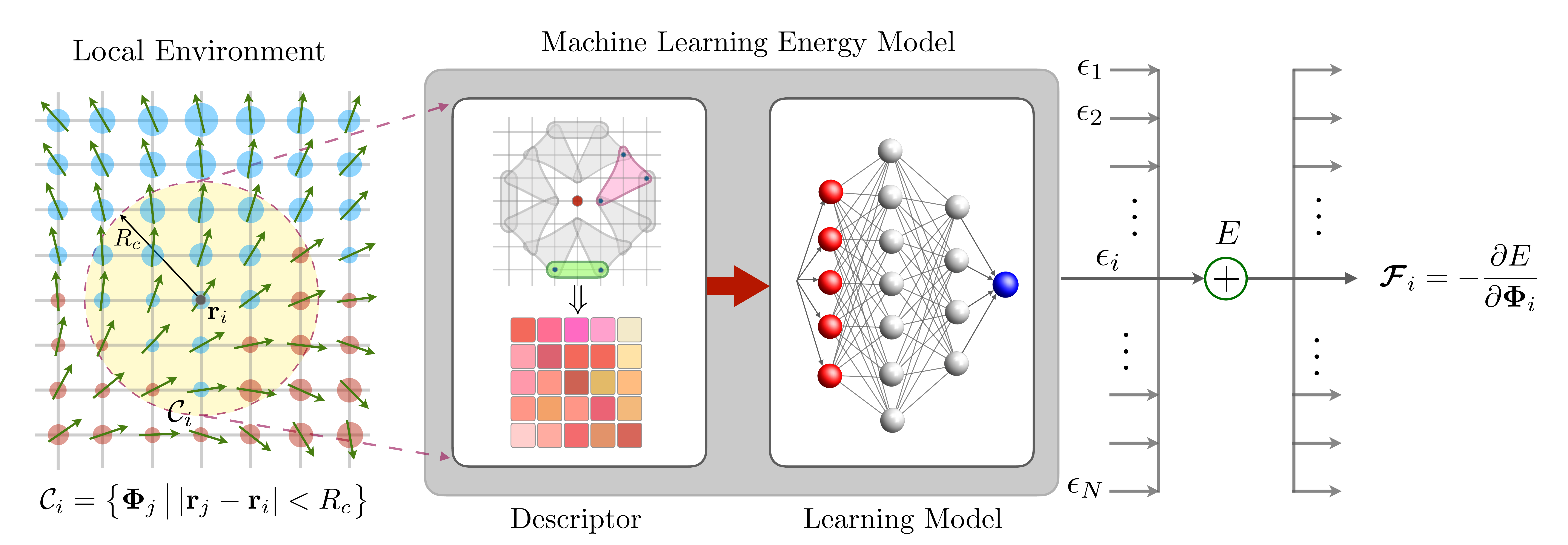}
\caption{Numerical framework of the machine-learning force-field model for dynamical simulations of condensed matter systems. At the center of this approach is the ML energy model which takes the local neighborhood $\mathcal{C}_i$ of a given site as the input, and predicts a local energy $\epsilon_i$. The generalized force $\bm{\mathcal{F}}_i$ is given by the derivative of the total energy $E = \sum_i \epsilon_i$. The ML energy model consists of two major and roughly independent components: the descriptor and the learning model. }
\label{fig:ml-potential} 
\end{figure*}

Compared with dynamical simulations based on an empirical energy model, the quantum dynamical approach here is to obtain the effective energy as well as forces by integrating out the electrons on the fly. 
However, the calculation of these effective forces is highly time-consuming, and could be prohibitively expensive for large systems.  For example, for lattice models without electron-electron interactions, i.e. $v = 0$, the calculation of the effective forces only requires diagonalizing a quadratic fermionic Hamiltonian. The time complexity of direct diagonalization for a system of $N$ sites scales as $\mathcal{O}(N^3)$. Even though linear-scaling techniques, such as the kernel polynomial method (KPM)~\cite{weisse06}, have been developed for computing the density matrices of such quadratic Hamiltonians, since the electronic problem has to be solved at every time-step of the dynamical simulation, sophisticated implementations, including for example GPU programming, are often required in order to meet the required efficiency. 

For models with electron interactions $v\neq 0$, such as the Hubbard-Kanamori-type interactions, more sophisticated many-body methods are needed to solve the lattice electron model. One popular and widely used approach is the self-consistent methods which include the well-known Hartree-Fock mean-field for symmetry-breaking phases and the Gutzwiller/slave-boson methods for Mott transitions. The central idea of this approach is to reduce the many-body problem into an effective single-particle or quadratic Hamiltonian, which can then be solved by either exact diagonalization or KPM. However, the requirement of self-consistency means that solution of the quadratic Hamiltonian has to be computed multiple times through iteration until a convergence is reached. And this iteration has to be performed again at every time-step, which introduces an extra time complexity even with efficient techniques such as the KPM. The computational overhead is even more demanding for more advanced methods such as the dynamical mean-field theory and quantum Monte Carlo simulations.

As discussed in Section~\ref{sec:intro}, ML methods offer a promising solution to this computationally difficult problem. The central idea is the principle of locality, also called the nearsightedness of electronic matter~\cite{kohn96,prodan05}, which assumes that local properties such as the on-site forces $\bm{\mathcal{F}}_i$ only depends on classical fields in the neighborhood of the $i$-th site. This approach is similar to the Behler-Parrinello (BP) formulation which is fundamental to the ML potential for {\em ab-initio} MD simulations~\cite{behler07,bartok10}.  Specifically, the total energy Eq.~(\ref{eq:E1}) is first partitioned into local energies associated with individual sites:
\begin{eqnarray}
	E = \sum_i \epsilon_i.
\end{eqnarray}
Importantly, the local site-energy $\epsilon_i$ is assumed to depend only on the classical fields in the local environment through a universal function $\epsilon_i = \varepsilon\bigl(\mathcal{C}_i \bigr)$, where $\mathcal{C}_i$ denotes the local configuration of the classical variables. In practical implementations, this is often defined as the $\bm\Phi_i$ variables with a given cutoff radius $R_c$: 
\begin{eqnarray}
	\label{eq:Ci}
	\mathcal{C}_i = \bigl\{ \bm \Phi_j \, \big| \, R_{ij} = |\mathbf r_j - \mathbf r_i | \le R_c \bigr\}.
\end{eqnarray}
With this partitioning, the calculation of the total electron energy can now be significantly simplified by properly grouping the classical variables $\bm\Phi_i$ and substituting into the universal function $\varepsilon(\mathcal{C}_i)$. Crucially, this complex function can now be accurately approximated by ML models, especially the deep-learning NN, thanks to their unprecedented expressive power. Practically, the ML model is derived through a training process based on solutions of the particular many-body method on small systems. Once this universal function is determined, the ML potential thus provides an effective  energy model in terms of the classical fields 
\begin{eqnarray}
	\label{eq:E-ML}
	E\bigl(\{\bm \Phi_i \} \bigr) = \sum_i \varepsilon(\mathcal{C}_i)
\end{eqnarray}
The effective forces Eq.~(\ref{eq:forces}) acting on the classical fields can now be efficiently computed from the derivatives of this classical energy. The general framework of the ML-based force field model for dynamical simulation is summarized in Fig.~\ref{fig:ml-potential}. For ML models based on the neural network, the forces can be readily obtained through the automatic differentiation.
Importantly, the ML model offers the efficiency of classical energy model, yet with the accuracy of the many-body techniques employed for generating the training dataset. 

\section{Descriptor}
\label{sec:descriptor}

The ML energy model in Eq.~(\ref{eq:E-ML}) naturally needs to preserve the symmetry of the original Hamiltonian, which includes both the symmetry of the dynamical variables and that of the underlying lattice. Nonetheless, despite the universal approximation capability of ML models, the symmetry of the original electron Hamiltonian is not automatically captured. Since the training of ML model is essentially an optimization  process with randomly chosen datasets, the symmetry of the model can only be statistically approximated even with a large amount of training data.  As discussed in Sec.~\ref{sec:intro}, a proper representation of the classical fields is required to ensure that the symmetry of the electron Hamiltonian is built into the ML model. A good representation, or descriptor, of the local environment must be invariant with respect to symmetry transformations of the system.

\begin{figure}
\includegraphics[width=1.0\columnwidth]{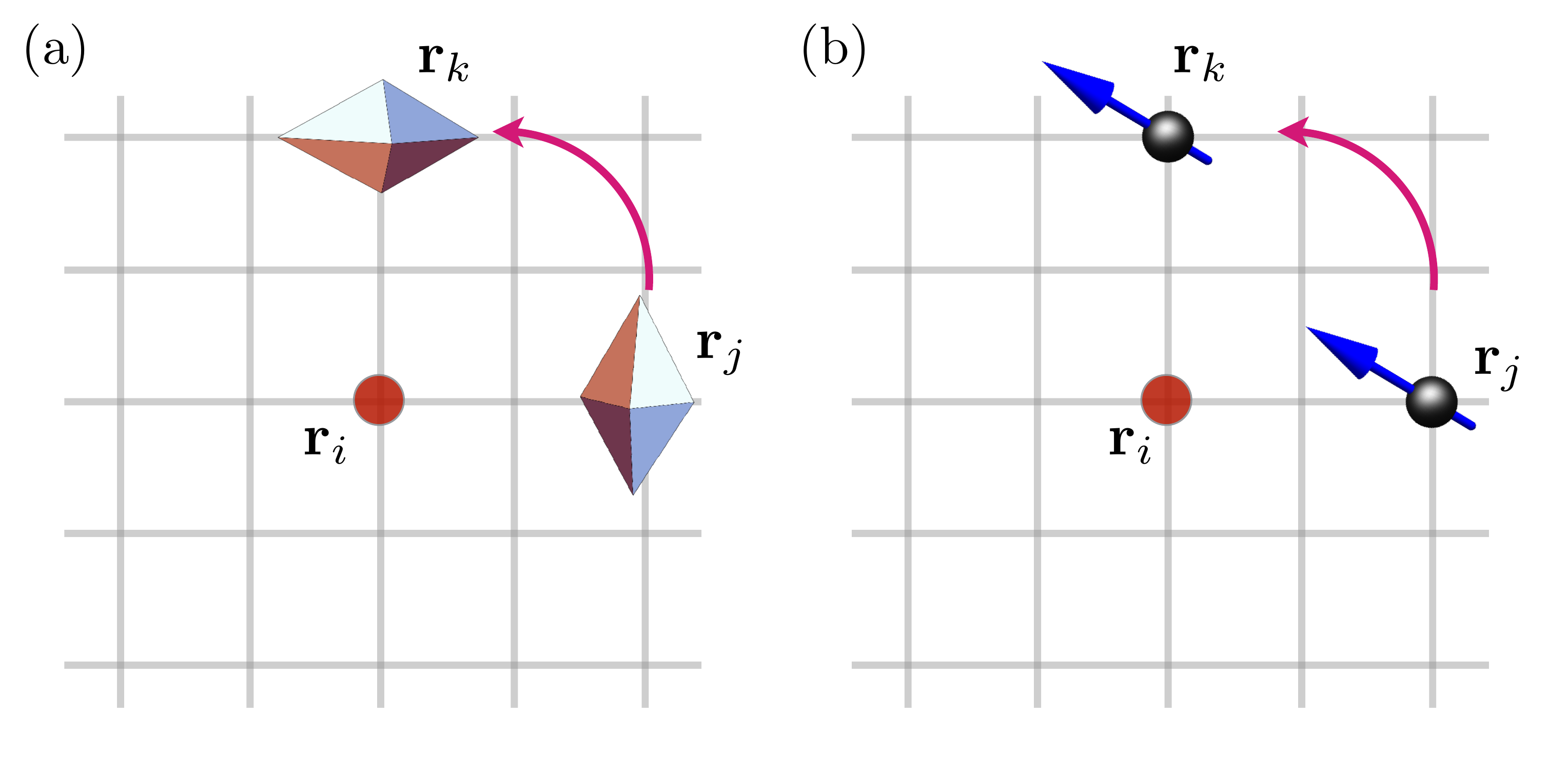}
\caption{Classical fields with different symmetry properties. An example of type-I case is the local Jahn-Teller distortion $\mathbf Q_j$ as shown in panel~(a). The lattice rotation/reflection is accompanied by a simultaneous transformation of the Jahn-Teller phonons. Panel~(b) shows the type-II case exemplified by local spins $\mathbf S_j$. The global rotation symmetry of spins is independent of the discrete point-group symmetry of the lattice.}
\label{fig:sym-types}  \
\end{figure}

For condensed matter systems defined on a lattice, the ML energy model $\varepsilon(\mathcal{C}_i)$ must be invariant under the discrete transformations of the point group, denoted as $G_L$, associated with the center site-$i$. Moreover, for classical fields with a complex structure, one also needs to take into account the symmetry associated with transformations among the multiple components $\Phi_{i, 1}, \Phi_{i, 2}, \cdots$ at the same site. We classify the classical fields into two types depending on whether these two symmetries are entangled to each other or not. Examples of these two types are illustrated in Fig.~\ref{fig:sym-types}. For models of the first type, the internal symmetry of the classical variables is coupled to the lattice symmetry. Examples of type-I classical variables include on-site displacement vector fields $\mathbf u_i = (u_i^x, u_i^y, u_i^z)$~\cite{mazumdar83,hirsch83,su80}, where the transformation of the 3 components of the displacement vector is coupled to the discrete rotations of the point group. Another example is the Jahn-Teller doublet $\mathbf Q_i = (Q_i^{x^2 - y^2}, Q_i^{3z^2 - r^2})$ characterizing local structural distortion~\cite{popovic00,sen06}. The only relevant symmetry group for such type-I models is the on-site point group $G_L$.  Under the symmetry operation $\hat{g} \in G_L$, the rearrangement of the classical fields at different lattice sites coincides with the transformation of the various components. Noting that $\Phi_{j, \alpha} = \Phi_\alpha(\mathbf r_j)$ with $\alpha = 1, 2, \cdots, M$ being the index of the various components, the transformation  of the classical fields is described by 
\begin{eqnarray}
	 \tilde\Phi_{\alpha}\bigl( O(\hat{g})\cdot \mathbf R_{ij} \bigr) = \mathcal{M}_{\alpha\beta}(\hat{g}) \Phi_{\beta}(\mathbf R_{ij}),
\end{eqnarray}
where $\mathbf R_{ij} = \mathbf r_j - \mathbf r_i$ is the relative position vector of site-$j$, $\mathcal{M}_{\alpha\beta}(g)$ is the $M$-dimensional matrix representation of the symmetry operation $\hat{g}$, and $O(\hat{g})$ is the 3-dimensional orthogonal matrix transforming site-$i$ to site-$k$, i.e. $\mathbf R_{ik}  = O(\hat{g}) \cdot \mathbf R_{ij} $.

For type-II models, the classical fields are characterized by an independent internal symmetry group, which will be denoted as $G_{\Phi}$. The most representative example, perhaps, is the models with local classical spins $\mathbf S_i$ as illustrated in Fig.~\ref{fig:sym-types}(b). For spins with $n$-component, the symmetry group of the system is a direct product of the lattice group $G_L$ and the internal symmetry group $G_{\Phi} = O(n)$ describing the global rotation symmetry of the spins. The most general symmetry operation consists of the lattice rotation/reflection $\hat{g} \in G_L$, and the transformation $\hat{h} \in G_{\Phi}$:
\begin{eqnarray}
	 \tilde\Phi_{\alpha}\bigl( O(\hat{g})\cdot \mathbf R_{ij} \bigr) = \mathcal{M}_{\alpha\beta}(\hat{h}) \Phi_{\beta}(\mathbf R_{ij}),
\end{eqnarray}
Note that $\mathcal{M}_{\alpha\beta}(\hat{h})$ is the matrix representation of the group $G_\Phi$, which is independent of the lattice point group. It is important to note that, for type-II models, the ML potential energy $\varepsilon(\mathcal{C}_i)$ must be invariant under the general combined symmetry transformation $\hat{h} \otimes \hat{g}$.

It is worth noting that while our focus is on the lattice models which are prevalent in condensed matter physics, most of the analysis presented in this work can be generalized to the off-lattice models or disordered systems if the point group $G_L$ is replaced by the continuous 3-dimensional rotation group $O(3)$, of course, assuming the system possesses such a global rotation symmetry. This means that our analysis can also be applied to electron models defined on an amorphous system or an atomic liquid state. In fact, the latter case can be viewed as a molecular dynamics system with an array of classical variables $\bm\Phi_i$ associated with every atom. A particular interesting application would be the Gutzwiller MD method where the classical fields $\bm\Phi_i$ corresponds to the slave-boson amplitudes~\cite{chern17}. 

Having discussed the general symmetry transformations for the two types of classical fields, we next describe a concise vector representation of the neighborhood $\mathcal{C}_i$ with its center at site-$i$, as defined in Eq.~(\ref{eq:Ci}).
We first consider the type-I models; the case of the type-II model will be discussed in Sec.~\ref{sec:type2}.  For convenience, the site-indices of lattice points within $\mathcal{C}_i$ are labeled as $j_{\textsf{r}}$, where $\textsf{r} = 1, 2, \cdots, L = |\mathcal{C}_i|$. Essentially, the integer $\textsf{r}$ offers an ordered list of lattice sites in the neighborhood.  Under the symmetry operation $\hat{g}$ of the point group, the lattice point $j_\textsf{r}$ is mapped to $j_\textsf{s}$ if and only if $(\mathbf r_{j_\textsf{s}} - \mathbf r_i) = O(\hat{g}) \cdot (\mathbf r_{j_\textsf{r}} - \mathbf r_i)$. Consequently, $\hat{g}$ can be represented by a $L \times L$ permutation matrix~$\mathcal{P}$, which means the nonzero matrix elements are $\mathcal{P}_{\textsf{s}\textsf{r}}(\hat{g}) = 1$ if the two sites $j_\textsf{r}$ and $j_\textsf{s}$ are related by $\hat{g}$. Next we introduce a vector $\vec{\mathcal{U}}$ whose components are given by the classical fields in the neighborhood:
\begin{eqnarray}
	\label{eq:U_def}
	\mathcal{U}_{\textsf{r},\alpha} = \Phi_\alpha(\mathbf r_{j_\textsf{r}}),
\end{eqnarray}
It is easy to see that $\vec{\mathcal{U}}$ offers a vector representation of dimension $L \times M$ for the point group $G_L$. And the corresponding matrix representation $\mathcal{T}$ of the symmetry operation $\hat{g} \in G_L$ is given by
\begin{eqnarray}
	\label{eq:transform-T}
	\tilde{\mathcal{U}}_{\textsf{r}, \alpha} = \mathcal{T}_{\textsf{r}\alpha, \textsf{s}\beta}(\hat{g}) \,\mathcal{U}_{\textsf{s},\beta} = \mathcal{P}_{\textsf{r}\textsf{s}}(\hat{g}) \mathcal{M}_{\alpha\beta}(\hat{g}) \,\mathcal{U}_{\textsf{s},\beta}.
\end{eqnarray}
Also importantly, the matrix $\mathcal{T}$ provides an orthogonal matrix representation of the point group, i.e. $\mathcal{T}^{\dagger} \mathcal{T} = \mathcal{T} \mathcal{T}^{\dagger } = \mathbb{I}$, where $\mathbb{I}$ is the $L\times M$-dimensional identity matrix.
Next we present two descriptors based on this vector representation of the neighborhood. 

\subsection{Correlation matrix}
\label{sec:corr-mat}

The idea of correlation matrix is similar to the so-called Weyl matrix~\cite{weyl46} for characterizing the local environment of a single-species molecular system. Specifically, for $L$ atoms within a cutoff radius in the neighborhood of atom-$i$, the Weyl matrix is defined as $\Sigma_{jk} = (\mathbf r_j - \mathbf r_i) \cdot (\mathbf r_k - \mathbf r_i)$. Since the matrix elements are given by scalar products of relative position vectors, the Weyl matrix remains the same under rotation, reflection, and translation operations. However, $\Sigma_{jk}$ is not a suitable descriptor because permutations of atoms change the order of rows and columns. On the other hand, such permutation operations correspond to a unitary or orthogonal transformation of the Weyl matrices. Consequently the eigenvalues $\{ \lambda_m \}$ of the $\Sigma$-matrix are invariant under permutation and can be used as a descriptor. A generalization of the Weyl matrix, which can also treat multiple atom-species, is the Coulomb matrix~\cite{rupp12}:  $M_{ij} = Z_i Z_j / |\mathbf r_i - \mathbf r_j|$ for $i\neq j$, where $Z_i$ is the nuclear charge of atom-$i$, and $M_{ii} = {\rm const} \times Z_i^{2.4}$. The ``Coulomb" interaction form of the off-diagonal matrix elements partly accounts for the Coulomb repulsion between the nuclei, which also highlights the importance of taking into account the atomic pair distances in the descriptor. Other forms of the pair correlation, such as Edward sum or sine-matrices, have also been proposed~\cite{himanen20}. 

Motivated by Weyl and Coulomb matrices, we propose a descriptor given by the ordered eigenvalues of the following correlation matrix of the classical fields
\begin{eqnarray}
	\label{eq:corr-m}
	C_{\textsf{r}\alpha, \textsf{s}\beta} = \left\{ \begin{array}{ll} g(\mathcal{U}_{\textsf{r}, \alpha}) \quad & (\textsf{r}\alpha) = (\textsf{s}\beta) \\ 
	f\bigl( \mathbf r_{j_\textsf{r}},  \mathbf r_{j_\textsf{s}} \bigr) \,\mathcal{U}_{\textsf{r}, \alpha} \, \mathcal{U}_{\textsf{s}, \beta} \quad & \mbox{otherwise} \end{array} \right. \quad
\end{eqnarray}
where $g(\cdot)$ and $f(\cdot)$ are two functions depending on the model under consideration. For example, one can choose a Coulomb interaction $f(\mathbf r_1, \mathbf r_2) = 1/|\mathbf r_1 - \mathbf r_2|$.  It is worth noting that this correlation matrix is {\em not} invariant under symmetry operations of the point group $G_L$. Instead, from Eq.~(\ref{eq:transform-T}), the $C$ matrix transforms as
\begin{eqnarray}
	  \tilde{C} = \mathcal{T}(\hat{g}) \, C \, \mathcal{T}^\dagger(\hat{g}).
\end{eqnarray}
Nonetheless, since $\mathcal{T}$ is an orthogonal matrix, the eigenvalues of the correlation matrix remains invariant with respect to symmetry operations $\hat{g}$ of the point group. As a result, the ordered list of eigenvalues $\lambda_m$ of the correlation matrix $C$ can be used as a descriptor that preserves the symmetry of the system. In practical implementations, only a finite number of the largest eigenvalues are used as feature variables. Unlike the Coulomb matrix used for fitting the atomization energies of molecular systems, there is no physical basis for the choice of the pair function $f(R)$.  An example of descriptor based on correlation matrix is given in Sec.~\ref{sec:scalar} below.

\subsection{Bispectrum coefficients}
\label{sec:bispectrum}

In this Section, we present a more systematic method for constructing a descriptor based on the group-theoretical method. Specifically, the feature variables are given by the so-called bispectrum coefficients computed from the expansion coefficients of irreducible representations of the point group~\cite{kakarala93,kondor07,kakarala09}. The bispectrum coefficients, which are invariant under the symmetry operations of the point group, are in a sense similar to the scalar triple product of three vectors which is invariant under arbitrary rotations.  It is also worth noting that similar group-theoretical methods, with important modifications to simplify the implementation, have been proposed as descriptor for ML interatomic potentials in quantum MD simulations~\cite{bartok10,bartok13}.

 In order to compute the bispectrum coefficients, the first step is to obtain the irreducible representations of the neighborhood. As discussed above, the vector $\vec{\mathcal{U}}$, defined in Eq.~(\ref{eq:U_def}) provides a $L\times M$-dimensional representation of the local environment $\mathcal{C}_i$. This high-dimensional representation can then be decomposed into irreducible representations (IRs) of the point group $G_L$ following the standard procedures~\cite{grouptheory,inui_book}. Specifically, we use $\Gamma$ to label the different IRs in the decomposition, and denote the corresponding basis vector of IR-$\Gamma$ as
 \begin{eqnarray}
 	\vec{\bm\Upsilon}^{ \Gamma} = \bigl(\vec{\Upsilon}^{\Gamma }_1, \vec{\Upsilon}^{\Gamma }_2, \cdots, \vec{\Upsilon}^{ \Gamma }_{n_\Gamma} \bigr),
 \end{eqnarray}
 where $n_\Gamma$ is the dimension of corresponding IR. Note that each ``component" $\vec{\Upsilon}^{\Gamma}_\mu = \{ \Upsilon^{\Gamma}_{\mu; \textsf{r}, \alpha} \}$ is itself a $(L\times M)$-dimensional vector.  The neighborhood vector is then decomposed as
 \begin{eqnarray}
 	\label{eq:decomp1}
 	\mathcal{U}_{\textsf{r}, \alpha} = \sum_{\Gamma} \sum_{\mu = 1}^{n_\Gamma} f^{\Gamma}_\mu \, \Upsilon^{\Gamma}_{\mu; \, \textsf{r}, \alpha}.
 \end{eqnarray}
 The expansion coefficients $f^{\Gamma}_\mu$ of the IR, play a role similar to the Fourier coefficients for the translation group. 
 Using the orthogonality of the basis vectors of different IRs, the expansion coefficients are given by
 \begin{eqnarray}
 	f^{\Gamma}_\mu = \vec{\Upsilon}^{\Gamma\dagger}_\mu \cdot \vec{\mathcal{U}} =  \sum_{\textsf{r} = 1}^{L} \sum_{ \alpha = 1}^M \Upsilon^{\Gamma*}_{\mu, \,\textsf{r} \alpha} \,\, \mathcal{U}^{\,}_{\textsf{r}\alpha}.
 \end{eqnarray}
 For convenience, we can group the expansion coefficients of a given IR into a vector: 
 \begin{eqnarray}
 	\bm f^{\Gamma} = \bigl( f^{\Gamma}_1, f^{\Gamma}_2, \cdots, f^{\Gamma}_{n_\Gamma} \bigr).
 \end{eqnarray}
 In terms of the classical fields, see e.g. Eq.~(\ref{eq:U_def}), the expansion coefficients are
 \begin{eqnarray}
	f^{\Gamma}_\mu = \sum_{\textsf{r} = 1}^{L} \sum_{ \alpha = 1}^M \Upsilon^{\Gamma*}_{\mu; \,\textsf{r} \alpha} \, \Phi_\alpha(\mathbf r_{j_{\textsf{r}}}).
 \end{eqnarray}
 Under symmetry operations $\hat{g}$ of the point group $G_L$, different IRs transform independently of each other. Consequently, the transformation of the vector $\bm f^{\Gamma}$ of a given IR is described by an $n_\Gamma \times n_\Gamma$ unitary matrix~$\bm D^{\Gamma}$ as
 \begin{eqnarray}
 	\label{eq:transform}
	\tilde{f}^{\,\Gamma}_\mu = \sum_{\mu'} D^{\Gamma}_{\mu  \mu'}(\hat{g})\, f^{\Gamma}_{\mu'},
\end{eqnarray} 
or the more concise vector equation: $\tilde{\bm f}^{\,\Gamma} = {\bm D}^{\Gamma}\cdot {\bm f}^{\Gamma}$.
From the transformation relation Eq.~(\ref{eq:transform-T}) for the vector $\vec{\mathcal{U}}$, the transformation matrix $\bm D^{\Gamma}$ can be explicitly computed. 
\begin{eqnarray}
	D^{\Gamma}_{\mu\mu'}(\hat{g}) = \vec{\Upsilon}^{\Gamma\dagger}_{\mu} \cdot \mathcal{T}(\hat{g}) \cdot \vec{\Upsilon}^{\Gamma}_{\mu'}.
\end{eqnarray}
It is worth noting that the transformation matrices of a given IR have been tabulated for most point and double groups. 
Similar to the ordinary Fourier analysis, we define the power spectrum for a given IR as
\begin{eqnarray}
	\label{eq:power1}
	p^{\Gamma} \equiv {\bm f}^{\Gamma\dagger} \cdot {\bm f}^{\Gamma} = \sum_{\mu=1}^{n_\Gamma} \left| f^{\Gamma}_\mu \right|^2
\end{eqnarray}
Since the transformation matrices are unitary ${\bm D}^\dagger {\bm D} = 1$, it is easy to see that the power spectrum is invariant under symmetry operations:
\begin{eqnarray}
	\tilde{p}^{\,\Gamma} = \tilde{\bm f}^{\,\Gamma \dagger} \cdot \tilde{\bm f}^{\,\Gamma} = {\bm f}^{\Gamma\dagger} {\bm D}^{\Gamma\dagger} {\bm D}^{\Gamma} {\bm f}^{\Gamma} 
	=  {\bm f}^{\Gamma\dagger} {\bm f}^{\Gamma} = p^{\Gamma}, \quad
\end{eqnarray}
This indicates that the amplitude of each IR can be used as the descriptor for the local environment $\mathcal{C}_i$.   However, the power spectrum $p^{\Gamma}$ is not a complete descriptor of the neighborhood function, since it neglects the weight distribution within each IR. Neither does it account for the relative phases between different IRs. 

A more complete description, which consists of a larger set of invariants, is given by the bispectrum of the IRs. To this end, we first consider the tensor product of coefficient vectors ${\bm f}^{\Gamma_1}\otimes {\bm f}^{\Gamma_2}$, which can be viewed as the expansion coefficients of the tensor-product $\vec{\mathcal{U}} \otimes \vec{\mathcal{U}}$ of the vector representation with a tensor-product basis $\vec{\Upsilon}^{\Gamma_1}_{\mu} \otimes \vec{\Upsilon}^{\Gamma_2}_{\nu}$. Under a symmetry operation, according to Eq.~(\ref{eq:transform}), the tensor-product transforms as
\begin{eqnarray}
	\label{eq:prod_f}
	& & {\bm f}^{\Gamma_1}\otimes {\bm f}^{\Gamma_2} 
	\to \left( {\bm D}^{\Gamma_1} \cdot {\bm f}^{\Gamma_1} \right) \! \otimes \! \left( {\bm D}^{\Gamma_2} \cdot {\bm f}^{\Gamma_2} \right) \nonumber \\
	& &  \qquad \qquad = \left({\bm D}^{\Gamma_1}\otimes {\bm D}^{\Gamma_2} \right)\cdot \left({\bm f}^{\Gamma_1}\otimes {\bm f}^{\Gamma_2} \right). 
\end{eqnarray}
As is well established in the representation theory of finite groups, the direct product of two IRs can be decomposed into a direct sum of IRs. This indicates the following decomposition of the direct-product matrices:
\begin{eqnarray}
	\label{eq:prod_D}
	{\bm D}^{\Gamma_1} \otimes {\bm D}^{\Gamma_2} = \bigl(\bm C^{\Gamma_1,\Gamma_2}\bigr)^\dagger \biggl[ \bigoplus_{\Gamma} {\bm D}^{\Gamma} \biggr] {\bm C}^{\Gamma_1, \Gamma_2}, \quad
\end{eqnarray}
where $\oplus$ means direct sum over the IRs of the direct product. We note that IR of the same dimension and symmetry could appear more than once in the direct sum.  The $\bm C^{\Gamma_1, \Gamma_2}$ is a unitary matrix of dimension $n_{\Gamma_1} \times n_{\Gamma_2}$; its matrix elements are known as the Clebsch-Gordan coefficients of the symmetry group under consideration. Explicitly, we have
\begin{eqnarray}
	& & D^{\Gamma_1}_{\mu \mu'}(\hat{g}) \, D^{\Gamma_2}_{\nu \nu'}(\hat{g})  \\
	& & \qquad  = \sum_{\Gamma } \sum_{\kappa, \kappa'} \bigl( C^{  \Gamma; \Gamma_1,\Gamma_2 }_{\kappa, \mu\nu} \bigr)^*  \, 
	D^{\Gamma}_{\kappa \kappa'}(\hat{g})  \,C^{ \Gamma; \Gamma_1,\Gamma_2 }_{\kappa', \mu'\nu'} \nonumber
\end{eqnarray}
As mentioned above, the sum over $\Gamma$ could include multiple IRs of the same transformation properties. 
To construct the bispectrum coefficients, we first consider the following vector:
\begin{eqnarray}
	\label{eq:V1}
	{\bm v}^{\Gamma_1, \Gamma_2} = {\bm C}^{\Gamma_1, \Gamma_2}\cdot ({\bm f}^{\Gamma_1} \otimes {\bm f}^{\Gamma_2} )
\end{eqnarray}
Since the Clebsch-Gordan matrix is essentially a transformation of basis, vector $\bm v$ is thus the expansion coefficients of the irreducible basis for the tensor-product $\vec{\mathcal{U}} \otimes \vec{\mathcal{U}}$. This can also be seen from the transformation of the $\bm v$ vector. Substitute Eq.~(\ref{eq:prod_D}) into (\ref{eq:prod_f}), and multiply the resultant expression by the ${\bm C}^{\Gamma_1, \Gamma_2}$ matrix from the left, we see that under symmetry operation $\hat{g}$, the vector ${\bm v}^{\Gamma_1, \Gamma_2}$ transforms according to
\begin{eqnarray}
	\tilde{\bm v}^{\Gamma_1, \Gamma_2} = \biggl[ \bigoplus_{\Gamma} {\bm D}^{\Gamma}(\hat{g}) \biggr] \cdot {\bm v}^{\Gamma_1, \Gamma_2}
\end{eqnarray}
This result thus also indicates we can decompose $\bm v$ into a direct sum of vectors each of which corresponds to an irreducible representation:
\begin{eqnarray}
	{\bm v}^{\Gamma_1, \Gamma_2} = \bigoplus_{\Gamma} \, {\bm u}^{\Gamma ; \Gamma_1, \Gamma_2}
\end{eqnarray}
Each vector transforms under symmetry operation as
\begin{eqnarray}
	\tilde{\bm u}^{\Gamma; \Gamma_1, \Gamma_2} = {\bm D}^{\Gamma}(\hat{g}) \cdot {\bm u}^{\Gamma; \Gamma_1, \Gamma_2}.
\end{eqnarray}
From this equation and Eq.~(\ref{eq:transform}) for the transformation of the vector $\bm f$ belong to the same IR-$\Gamma$, it is straightforward to see that the following ``inner product" is a scalar invariant under any symmetry operation:
\begin{eqnarray}
	b^{\Gamma, \Gamma_1, \Gamma_2}  = {\bm f}^{\Gamma \,\dagger} \cdot {\bm u}^{\Gamma; \Gamma_1,   \Gamma_2},
\end{eqnarray}
These coefficients are called the bispectrum of the expansion coefficients of the IRs. Using Eq.~(\ref{eq:V1}) to express the $\bm u$ vectors, we obtain the following explicit formula for the bispectrum coefficients 
\begin{eqnarray}
	\label{eq:bispectrum1}
	b^{\Gamma, \Gamma_1, \Gamma_2}   = \sum_{\kappa,\mu,\nu} 
	C^{\Gamma; \Gamma_1, \Gamma_2}_{\kappa, \mu\nu} f^{\Gamma *}_\kappa f^{\Gamma_1}_\mu f^{\Gamma_2}_\nu. \quad 
\end{eqnarray}
The above expression shows the similarity of the $b$ coefficients with the scalar triple of three O(3) vectors. It should also be noted that the power spectrum $p^{\Gamma}$ is part of the bispectrum coefficients. In fact, while formally the bispectrum coefficients are built from product of three IR-amplitudes, they can also be used to describe invariants consisting of two IR-coefficients. This corresponds to the case when the decomposition of the direct product representation $\Gamma_1 \otimes \Gamma_2$ includes the trivial one-dimensional representation, denoted as $\Gamma_0$ for convenience. By setting the corresponding coefficient to be a constant, e.g. $f_{\Gamma_0} = 1$, we see that the resultant bispectrum coefficient $b^{\Gamma_0, \Gamma_1, \Gamma_2}$ is nonzero only if the two IRs $\Gamma_1$ and $\Gamma_2$ transform in exactly the same way under symmetry operations, hence have the same dimension. Consequently, we can define the following generalization of power spectrum
\begin{eqnarray}
	\label{eq:general_p}
	p^{\Gamma_1, \Gamma_2} = {\bm f}^{\Gamma_1 \dagger} \cdot {\bm f}^{\Gamma_2} =  \sum_{\mu}  f^{\Gamma_1 *}_\mu f^{\Gamma_2}_\mu.
\end{eqnarray}  
The standard power spectrum  Eq.~(\ref{eq:power1}) of a given IR-$\Gamma$  corresponds to the case $\Gamma_1 = \Gamma_2 = \Gamma$.

Importantly, since the bispectrum coefficients are invariant under symmetry operations of the point group, they serve as proper descriptor to be combined with the ML models. Moreover, it can be shown that the bispectrum provides a faithful representation of the original configuration in the sense that the vector $\mathcal{U}$ can be rigorously reconstructed from all bispectrum coefficients~\cite{kakarala93,kondor07,kakarala09}.  For practical applications, however, there are a large number of the bispectrum coefficients for most models and point groups. For example, let $\mathbb{N}$ be the number of IRs from the decomposition of $\vec{\mathcal{U}}$, which is roughly of the order of $\mathbb{N} \sim (L \times M)$, the number of bispectrum is of the order of~$\mathbb{N}^3$, which in general is a rather large number.  Moreover, as will be demonstrated in explicit examples in Sec.~\ref{sec:scalar}, the bispectrum is an over-complete representation with redundant information. Consequently, further simplification is often required for practical implementations.

As an application of the bispectrum method, here we briefly review its application to represent the atomic environment for ML interatomic potentials. The bispectrum method is often combined with the Gaussian kernel potential learning model and the so-called smooth overlap of atomic positions (SOAP) technique, which approximates atoms in the neighborhood by Gaussian functions of a finite width~\cite{bartok10,bartok13}. For MD simulations, the local atomic configuration is described by the charge density $\rho(\mathbf r)$ with the origin $\mathbf r= 0$ corresponding to the center atom. The symmetry group of three-dimensional free space is $G_L$ = SO(3), and the corresponding irreducible representations are labeled by an integer  $\ell = 0, 1, 2, \cdots$, which is essentially the angular momentum quantum numbers~\cite{qm_angular}. Indeed, the basis function $\Upsilon^{\Gamma}_\mu$ for the SO(3) group is simply the spherical harmonics $Y_{\ell, m}$. Choosing a proper radial basis $g_n(r)$, the atomic neighborhood density is expanded as
\begin{eqnarray}
	\rho(\mathbf r) = \sum_{n=0}^\infty \sum_{\ell = 0}^\infty \sum_{m = -\ell}^{\ell} f_{n\ell m} \, g_n(r) Y_{\ell m}(\theta, \phi),
\end{eqnarray}
Note that there is an additional integer index $n$ for the expansion coefficients due to the radial dependence. The bispectrum coefficients are then labeled by six integers~\cite{bartok13}:
\begin{eqnarray}
	b^{\ell; \ell_1, \ell_2}_{n; n_1, n_2} = \sum_{m, m_1, m_2} f^*_{n \ell m} C^{\ell; \ell_1, \ell_2}_{m; m_1, m_2} \,f^{\,}_{n_1 \ell_1 m_1} f^{\,}_{n_2 \ell_2 m_2}, \qquad
\end{eqnarray}
where $C^{\ell; \ell_1, \ell_2}_{m; m_1, m_2}$  are Clebsch-Gordan coefficients of the SO(3) group~\cite{qm_angular}. For a given set of radial indices $(n, n_1, n_2)$, the bispectrum coefficients are nonzero only when $\ell = \ell_1 + \ell_2$ due to conservation of angular momentum. However, there are still an infinite number of the $b$ coefficients, and some cutoff $\ell_{\rm max}$ has to be introduced for practical implementation.  To further simplify the calculation, one can consider only coefficients with $n_1 = n_2 = n$. This, however, implies that rotations of different radial basis are decoupled, thus introducing a spurious symmetry. Nonetheless, some simplifications can be achieved through special designs of the radial basis functions~\cite{bartok13}.

Instead of dealing with the natural SO(3) group for the three-dimensional space, an alternative approach is to project the atomic environment within a cutoff $R_c$ onto the surface of the four-dimensional sphere $S^3$~\cite{bartok10,bartok13}. Specifically, this means that the center-atom is at the north pole, while the cutoff radius, i.e. the 3-sphere specified by $|\mathbf r| = R_c$, is mapped to the south pole of the $S^3$. Next assuming an approximate SO(4) symmetry for the projected atomic density, one can then use the resultant bispectrum coefficients as the descriptor. As the IR of the SO(4) group is again labeled by an integer $j$, the bispectrum coefficients are indexed by three integers $b^{j, j_1, j_2}$. It should be noted that although the projection to $S^3$ implicitly assumes a spurious SO(4) symmetry, a most crucial advantage of this approach is the absence of the need for radial basis.

\subsection{Internal symmetry}
\label{sec:type2}

As discussed above, the type-II models are characterized by an internal symmetry group $G_{\Phi}$, independent of the lattice point group, that governs the transformation of the classical fields $\bm\Phi_i$. The feature variables for the ML models need to be invariant with respect to transformations of both symmetry groups. As the multiple components of the local classical vector $\bm\Phi_i$ do not transform simultaneously with the lattice symmetry operations, the method described in Sec.~\ref{sec:bispectrum} cannot be directly applied to the type-II models. 

One solution is to treat each of the $M$ components of the classical fields $\bm\Phi_i = \{ \Phi_{i, \alpha} \}$ ($\alpha = 1, 2, \cdots, M$) as independent. We then view the neighborhood configuration $\vec{\mathcal{U}}_{\alpha} = (\mathcal{U}_{1, \alpha}, \mathcal{U}_{2, \alpha}, \cdots, \mathcal{U}_{L, \alpha} )$ as $M$  independent $L$-dimensional representations of the neighborhood $\mathcal{C}_i$.
Each component is then decomposed into the IRs of the lattice group (c.f. Eq.~(\ref{eq:decomp1}) for the type-I case)
\begin{eqnarray}
	\mathcal{U}_{\textsf{r}, \alpha} = \sum_\Gamma \sum_{\mu = 1}^{n_\Gamma} f^{\Gamma}_{\mu, \alpha} \, \Upsilon^{\Gamma}_{\mu; \textsf{r}},
\end{eqnarray}
Note the basis function $\Upsilon$ of the IR now only depends on the site-index $\textsf{r}$. The coefficients of the IRs are similarly obtained based on the orthogonality of the basis functions
\begin{eqnarray}
	\label{eq:f_Phi2}
	f^{\Gamma}_{\mu, \alpha} = \sum_{\textsf{r}=1}^L \Upsilon^{\Gamma *}_{\mu; \textsf{r}} \, \mathcal{U}_{\textsf{r}, \alpha} 
	=  \sum_{\textsf{r}=1}^L \Upsilon^{\Gamma *}_{\mu; \textsf{r}} \,\Phi_\alpha(\mathbf r_{j_{\textsf{r}}} ). 
\end{eqnarray}
For each of the IR $\Gamma$ in the decomposition (with respect to point group), there are $M$ components indexed by $\alpha$. As each can be viewed as a $M$-dimensional representation of the internal symmetry group, it can be decomposed into the IR of $G_\Phi$ labeled by $\textsf{K}$:
\begin{eqnarray}
	f^\Gamma_{\mu, \alpha} =  \sum_{\textsf{K}}\sum_{\textsf{m} = 1}^{n_{\textsf{K}}} {F}^{ \Gamma, \textsf{K}}_{\mu, \textsf{m}} \, \mathcal{Y}^{\textsf{K}}_{\textsf{m},  \alpha}.
\end{eqnarray}
Here $\mathcal{Y}^{\textsf{K}}_{\textsf{m}}$ is the basis function of the $\textsf{K}$-th IR whose dimension is $n_{\textsf{K}}$. Using the orthogonality of the basis functions, the expansion coefficients are given by
\begin{eqnarray}
	F^{\Gamma, \textsf{K}}_{\mu , \textsf{m}} = \sum_{\alpha = 1}^M \mathcal{Y}^{\textsf{K} *}_{\textsf{m}, \alpha} \, f^{\Gamma}_{\mu, \alpha}
	= \sum_{\alpha = 1}^M \sum_{\textsf{r}=1}^L  \mathcal{Y}^{\textsf{K} *}_{\textsf{m}, \alpha} \, \Upsilon^{\Gamma *}_{\mu; \textsf{r}} \,\Phi_\alpha(\mathbf r_{j_{\textsf{r}}} ). \quad
\end{eqnarray}
Here we have used Eq.~(\ref{eq:f_Phi2}) in the second equality to express ${F}^{\Gamma, \textsf{K}}_{\mu , \textsf{m}}$ in terms of the classical fields. It is worth noting that this mixed expansion coefficients, expressed as a special combination of the classical fields, have well defined transformation properties, indicated by the IR indices $\Gamma$ and $\textsf{K}$, under both the point group of the site-symmetry and the internal symmetry group. However, since the two set of symmetry transformations are independent of each other, one cannot obtain simultaneous bispectrum coefficients with respect to both symmetry groups. To proceed, we can first ``trace out" the point group indices $\mu$ by forming the bispectrum coefficients of the point group first
\begin{eqnarray}
	B^{\Gamma, \Gamma_1, \Gamma_2}_{\textsf{K}_1, \textsf{l}; \textsf{K}_2, \textsf{m}; \textsf{K}_3, \textsf{n}}   = \sum_{\kappa,\mu,\nu} 
	C^{\Gamma; \Gamma_1, \Gamma_2}_{\kappa, \mu\nu} F^{\Gamma, \textsf{K}_1 *}_{\kappa, \textsf{l}} F^{\Gamma_1, \textsf{K}_2}_{\mu, \textsf{m}} F^{\Gamma_2, \textsf{K}_3}_{\nu, \textsf{n}}. \quad 
\end{eqnarray} 
These coefficients with three indices $\textsf{l}$, $\textsf{m}$, $\textsf{n}$ can be viewed as a tensor-product representation $\textsf{K} \otimes \textsf{K}_1 \otimes \textsf{K}_2$ of the internal symmetry group $G_\Phi$.   Next we decompose this tensor-product representation into a direct sum of IRs of the group $G_\Phi$. For convenience of the discussion, we denote the coefficients of the IR in the direct sum as $\texttt{F}^\textsf{K}_\textsf{q}$. Then invariants with respect to the internal symmetry are given by the bispectrum coefficients from the ``triple" product of these $\texttt{F}^\textsf{K}_\textsf{q}$ coefficients. Importantly, these bispectrum coefficients are now invariant with respect to both the lattice and internal symmetry groups. Since the $\texttt{F}^\textsf{K}_\textsf{q}$ coefficients themselves are already triple product of the field variables, the final invariants in general are composed of 9 classical variables; although some of them can be reduced. Since the number of the coefficients increases even more dramatically with the cutoff radius $R_c$ for type-II models, further approximations are necessary to simplify the implementation of the descriptor.  

A second approach, which is physically more intuitive and transparent, is to start from the symmetry of the classical fields and  first construct building blocks that are already invariant under the transformations of the internal symmetry group. The group-theoretical method discussed in Sec.~\ref{sec:bispectrum} is then applied to these building blocks for the lattice symmetry. To this end, we again note that the classical fields $\bm\Phi_j = \{ \Phi_{j, \alpha} \}$ at every sites in the neighborhood $\mathcal{C}_i$ is obviously an $M$-dimensional representation of the internal symmetry group, and can be decomposed into IRs of the $G_\Phi$ group:
\begin{eqnarray}
	\label{eq:decomp3}
	\Phi_{j, \alpha} = \sum_{\textsf{K}} \sum_{\textsf{m} = 1}^{n_{\textsf{K}}} \textsf{f}^{\,\textsf{K}}_{j,  \textsf{m}}  \, \mathcal{Y}^{\textsf{K}}_{\textsf{m}, \, \alpha},
\end{eqnarray}
It is worth noting that the expansion coefficients $\textsf{f}^{\,\textsf{K}}_{j, \textsf{m}}$ acquires a site index $j$. Again, using the orthogonality of $\mathcal{Y}$, we have
\begin{eqnarray}
	\textsf{f}^{\,\textsf{K}}_{j, \textsf{m}} = \sum_{\alpha = 1}^M \mathcal{Y}^{\textsf{K} \,*}_{\textsf{m}, \, \alpha} \, \Phi^{\,}_{j, \alpha}.
\end{eqnarray}
If the decomposition in Eq.~(\ref{eq:decomp3}) includes the  trivial representation $\textsf{K}_0$ which is by definition a one-dimensional IR, then the coefficients $\textsf{f}^{\,\textsf{K}_0}_j$ are automatically invariant with respect to the internal symmetry group and are part of the building blocks for the lattice group.

\begin{figure}
\includegraphics[width=1.0\columnwidth]{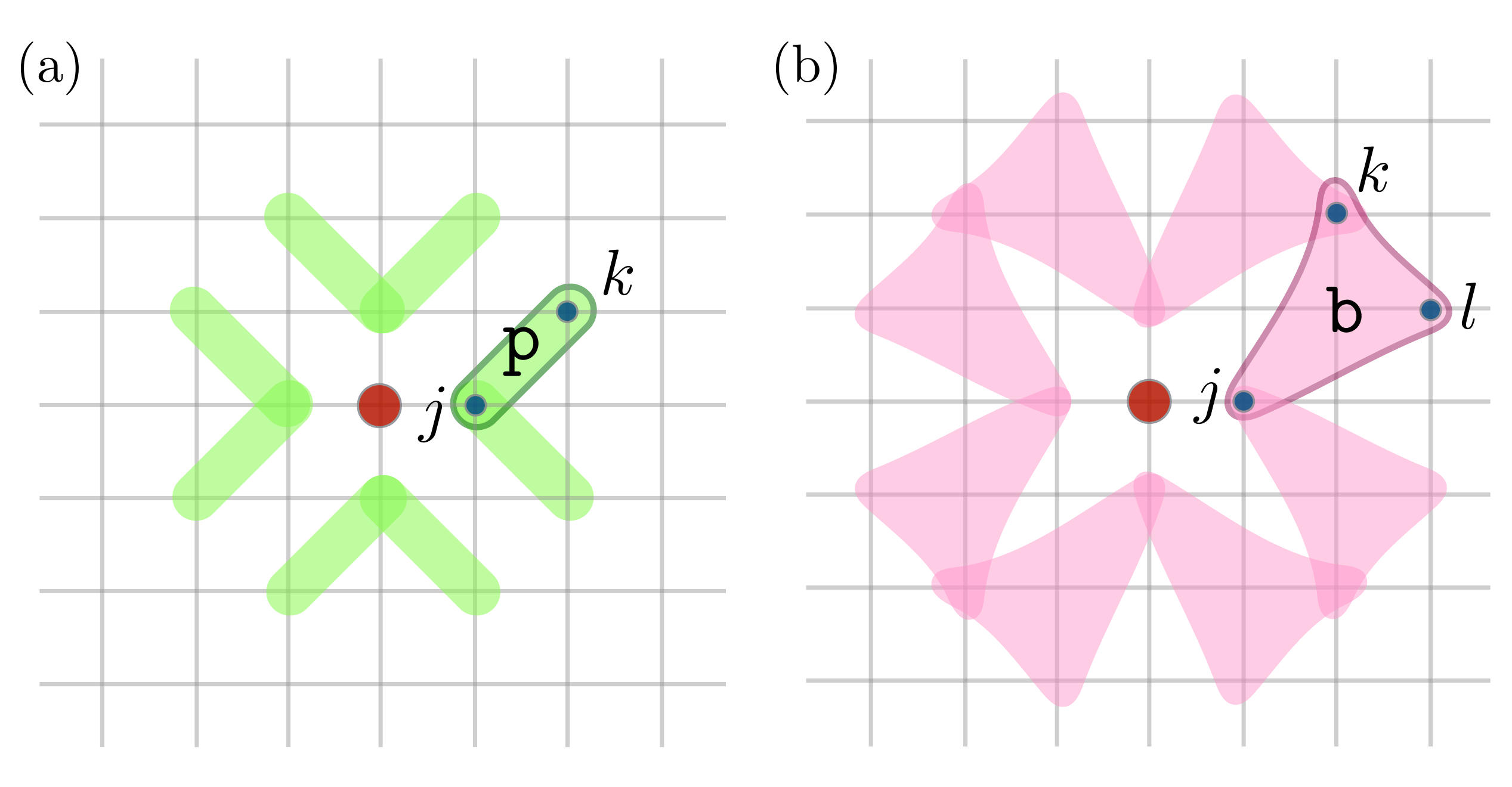}
\caption{Examples showing the atomic pair $(jk)$ and triplet $(jkl)$, which are related by the lattice rotation and reflection symmetries, in the neighborhood of the center site on a square lattice.}
\label{fig:blocks}  \
\end{figure}

Other invariants of the internal symmetry group are provided by the generalized power spectrum Eq.~(\ref{eq:general_p}) and the bispectrum coefficients. The crucial difference here is that these invariants are to be built from different sites, thus also serving as many-body correlation functions. First, we consider the generalized power spectrum obtained from a pair of sites $(jk)$
\begin{eqnarray}
	\label{eq:p-block}
	\texttt{p}^{\textsf{K}_1, \textsf{K}_2}_{jk} = \sum_{\textsf{m}} 
	\textsf{f}^{\,\textsf{K}_1 \, *}_{j, \textsf{m}} \, \textsf{f}^{\,\textsf{K}_2}_{k, \textsf{m}},
\end{eqnarray}
Again, the generalized power spectrum coefficient is nonzero only if the two IRs $\textsf{K}_1$ and $\textsf{K}_2$ have the same transformation properties.
 Similarly, one can build invariants of internal symmetry from a triplet $(jkl)$ of lattice sites based on the bispectrum coefficients
\begin{eqnarray}
	\label{eq:b-block}
	\texttt{b}^{\textsf{K}, \textsf{K}_1, \textsf{K}_2}_{jkl} = \sum_{\textsf{l}, \textsf{m}, \textsf{n}}  \texttt{C}^{\textsf{K}; \textsf{K}_1, \textsf{K}_2}_{\textsf{l}; \textsf{m}, \textsf{n}} \,
	\textsf{f}^{\,\textsf{K} \, *}_{j, \textsf{l}} \textsf{f}^{\,\textsf{K}_1}_{k, \textsf{m}} \textsf{f}^{\,\textsf{K}_2}_{l, \textsf{n}}.
\end{eqnarray}
where $ \texttt{C}^{\textsf{K}; \textsf{K}_1, \textsf{K}_2}_{\textsf{l}, \,\textsf{m}, \textsf{n}} $ are the Clebsch-Gordan coefficients of the internal symmetry group $G_\Phi$.
Fig.~\ref{fig:blocks} shows examples of the atomic pairs $(jk)$ and triplets $(jkl)$ related by the lattice rotation and reflection in the neighborhood of the center site. As mentioned above, these quantities $\texttt{p}$ and $\texttt{b}$ also encode the two-body and three-body correlations, respectively, of the neighborhood. Also importantly, they remain unchanged under operations of the internal symmetry group and can be used as building blocks for constructing the invariants of the lattice point group. To this end, we arrange them, including the single-site trivial IR, into a vector of dimension $\mathcal{N}$:
\begin{eqnarray}
	\label{eq:U3}
	\vec{\mathcal{U}} = ( \mathcal{U}_1, \mathcal{U}_2, \cdots, \mathcal{U}_{\mathcal{N}} ) 
	= \bigl( \textsf{f}^{\,\textsf{K}_0}_j, \, \texttt{p}^{\textsf{K}_1, \textsf{K}_2}_{jk}, \, \texttt{b}^{\textsf{K}, \textsf{K}_1, \textsf{K}_2}_{jkl} \bigr). \quad
\end{eqnarray}
Here we use $\mathcal{U}_J$ to denote the components of this vector, where the index $J$ is used to label either a site $j$, a pair $(jk)$, or a triplet $(jkl)$. The dimension $\mathcal{N}$ is dominated by the number of atomic pairs and triplets in the neighborhood.  For a neighborhood consisting of $L$ sites, these two number  scale as $L^2$ and~$L^3$, respectively. Moreover, one also needs to take into account the number of different IRs. As the total classical degrees of freedom is $L \times M$, the set of all $\texttt{f}$, $\texttt{p}$, and $\texttt{b}$ invariants obviously is an over-complete representation of the neighborhood. Practically, one needs to introduce further constraints in order to reduce this number, for example, by restricting distances between the pairs or triplets to be smaller than another cutoff, or to avoid too many overlaps of the pairs and triples.

Irrespective of the approximations, by keeping all symmetry related pairs and triples, as shown in Fig.~\ref{fig:blocks}, in Eq.~(\ref{eq:U3}),  the vector $\vec{\mathcal{U}}$ forms an $\mathcal{N}$-dimensional representation of the lattice point group $G_L$. We next apply the same group-theoretical method discussed in Sec.~\ref{sec:bispectrum} to obtain the bispectrum coefficients of the point group. We again decompose $\vec{\mathcal{U}}$ into the  IRs
\begin{eqnarray}
	\mathcal{U}_J = \sum_\Gamma \sum_{\mu=1}^{n_\Gamma} f^{\Gamma}_{\mu} \Upsilon^{\Gamma}_{\mu;  J}. 
\end{eqnarray}
where $\Upsilon^{\Gamma}_{\mu; J}$ are the appropriate basis functions. It is worth noting that the IRs of the single sites, pairs, and triplets are decoupled from each other. The expansion coefficients are then obtained separately as
\begin{eqnarray}
	f^{\Gamma}_\mu = \left\{ \begin{array}{l} \sum_{j\,\,\,} \Upsilon^{\Gamma *}_{\mu; j} \, \texttt{f}^{\,\textsf{K}_0}_{j} \\ \\
	\sum_{(jk) \,} \Upsilon^{\Gamma *}_{\mu; jk} \, \texttt{p}^{\textsf{K}_1, \textsf{K}_2}_{jk} \\ \\
	\sum_{(jkl)} \Upsilon^{\Gamma *}_{\mu; jkl} \, \texttt{b}^{\textsf{K}, \textsf{K}_1, \textsf{K}_2}_{jkl}  \end{array} \right.
\end{eqnarray}
Given these IR coefficients, Eqs.~(\ref{eq:bispectrum1}) and (\ref{eq:general_p}) can then be used to compute the generalized power spectrum and bispectrum coefficients, respectively, which are invariant with respect to both the internal and the lattice symmetry groups of the type-II systems. 

\subsection{Atom-centered symmetry functions}
\label{sec:scsf}

The building blocks introduced in Eqs.~(\ref{eq:p-block}) and~(\ref{eq:b-block}) above also offer the basis for a descriptor which can be viewed as the generalization of the atom-centered symmetry function (ACSF) originally proposed to describe the atomic configurations~\cite{behler07,behler16}. Unlike the group-theoretic methods, the ACSF approach is physically more intuitive and relatively simple to implement. On the other hand, it is more difficult to control the errors due to the {\em ad hoc} parameterizations of the symmetry functions. Nonetheless, ACSF has been successfully applied to the ML interatomic potential for a wide range of materials. We first briefly review the basic features of ACSF using the example of mono-atomic systems. For a given atomic configuration $\{\mathbf r_j\}$ in the vicinity of a center atom-$i$, the fundamental invariants that are invariant under rotations and reflections of the O(3) group are the distances $R_{ij} = | \mathbf r_j - \mathbf r_i|$ from the center atom, and the angles $\theta_{ijk} = \arccos[(\mathbf r_j - \mathbf r_i)\cdot (\mathbf r_k - \mathbf r_i) / R_{ij} R_{ik} ]$. Based on these quantities, two kinds of symmetry functions are introduced. The first type is the two-body (between atoms~$j$ and the center atom-$i$) symmetry function
\begin{eqnarray}
	G_2(\{\xi_m\}) = \sum_{j \neq i} F_2(R_{ij}; \{\xi_m\}),
\end{eqnarray}
where $F_2(R; \xi_m)$ is a user-defined function, parameterized by $\{\xi_m\}$ to extract atomic structures at certain distances from the center atom. One popular choice, proposed in the original work~\cite{behler07}, is a Gaussian with a soft cutoff at radius $R_c$
\begin{eqnarray}
	\label{eq:F2-envelop}
	F_2(R; \{\xi_m\}) = e^{- ( R - \xi_1)^2/\xi_2^2 }\, f_c(R ).
\end{eqnarray}
Here $f_c(r) = \frac{1}{2} \bigl[ \cos(\frac{\pi r}{R_c}) + 1 \bigr]$ for $R \le R_c$ and zero otherwise. The two parameters $\xi_1$ and $\xi_2$ speficiy the center and width, respectively, of the Gaussian function. The 3-body symmetry functions are defined as
\begin{eqnarray}
	 G_3(\{\xi_m\}) &=& \sum_{j, k \neq i} F_3(R_{ij}, R_{ik}, R_{jk}, \theta_{ijk}; \{\xi_m\}),
\end{eqnarray}
An example of the three-body envelop function characterized by three parameters is~\cite{behler07,behler16} 
\begin{eqnarray}
	\label{eq:F3-envelop}
	& & F_3(R_1, R_2, R_3, \theta; \{\xi_m\}) =  2^{1- \xi_1}  (1 + \xi_2 \cos\theta )^{\xi_1}  \\
	& & \qquad \times \exp\bigl[-( R_1^2 + R_1^2 + R_3^2)/\xi_3^2 \bigr] f_c(R_1) f_c(R_2) f_c(R_3). \nonumber
\end{eqnarray}
We note that generalizations to take into account the different atom species have also been made~\cite{himanen20}. Moreover, depending on the problems at hand, it might be more convenient to use different $F_2$ and $F_3$ functions, and several variants of these functions have been proposed~\cite{himanen20}.

Next we present a generalization of the ACSF for condensed-matter systems, where each atom is now associated with a dynamical classical field $\bm\Phi_j$. We emphasize that the formulation presented here can also be used for disordered systems, where the ``lattice" point group is replaced by the 3D rotation group SO(3). Moreover, for applications to MD simulation of liquid systems with a dynamical classical fields, the generalized ACSF provides a convenient descriptor for ML energy models for both the atomic dynamics and the classical fields. In order to incorporate the internal symmetry, our approach is to define a set of symmetry functions based on the building blocks in Eq.~(\ref{eq:U3}). We start with the two-body symmetry functions that include the coefficients of the trivial IR at every sites: 
\begin{eqnarray}
	\label{eq:G2a}
	G_{2a}(\{\xi_m\}) = \sum_{j \neq i} \texttt{f}^{\textsf{K}_0}_j F_2(R_{ij}; \{\xi_m\}) ,
\end{eqnarray}
This is the direct generalization of the original two-body symmetry functions that incorporates the on-site classical fields. Another way to build the 2-body symmetry functions is to use the invariants $\texttt{p}^{\textsf{K}_1, \textsf{K}_2}_{ij}$ between the center site-$i$ and a neighboring site-$j$:
\begin{eqnarray}
	\label{eq:G2b}
	G^{\textsf{K}_1, \textsf{K}_2}_{2b}(\{\xi_m\}) = \sum_{j \neq i} \texttt{p}^{\textsf{K}_1, \textsf{K}_2}_{jk} \, F'_2(R_{ij}; \{ \xi_m \}) , 
\end{eqnarray}
The envelope function $F_2'(R)$ is not necessarily the same as the one for $G_{2a}$. A three-body symmetry function based on single-site invariants is
\begin{eqnarray}
	\label{eq:G3a}
	& & G^{\textsf{K}_1, \textsf{K}_2}_{3a}(\{\xi_m\}) = \sum_{jk \neq i} \texttt{f}^{\textsf{K}_1}_j \texttt{f}^{ \textsf{K}_2}_{k} \nonumber \\
	& & \qquad \quad \times F^{\,}_3(R_{ij}, R_{ik}, R_{jk}, \theta_{ijk}; \{\xi_m\}),  
\end{eqnarray}
 The pair-wise invariants can also be combined with the center atom to define a three-body symmetry function:
\begin{eqnarray}
	\label{eq:G3b}
	& & G^{\textsf{K}_1, \textsf{K}_2}_{3b}(\{\xi_m\}) = \sum_{jk \neq i} \texttt{p}^{\textsf{K}_1, \textsf{K}_2}_{jk} \nonumber \\ 
	& & \qquad \quad \times F'_3(R_{ij}, R_{ik}, R_{jk}, \theta_{ijk}; \{\xi_m\}),  
\end{eqnarray}
A second type of 3-body symmetry functions is obtained from the invariants $\texttt{b}^{\textsf{K}, \textsf{K}_1, \textsf{K}_2}_{ijk}$ that involves the center atom
\begin{eqnarray}
	\label{eq:G3c}
	& & G^{\textsf{K}, \textsf{K}_1, \textsf{K}_2}_{3c}(\{\xi_m\}) = \sum_{jk \neq i} \texttt{b}^{\textsf{K}, \textsf{K}_1, \textsf{K}_2}_{ijk} \nonumber \\
	& & \qquad \quad \times F''_3(R_{ij}, R_{ik}, R_{jk}, \theta_{ijk}; \{\xi_m\}) ,  
\end{eqnarray}
Finally, several four-body symmetry functions can be defined based on the fundamental invariants of the internal symmetry group. For example, combining the triplet $(jkl)$ with the center site, we have 
\begin{eqnarray}
	\label{eq:G4}
	& & G^{\textsf{K}, \textsf{K}_1, \textsf{K}_2}_{4}(\{\xi_m \}) = \sum_{jkl \neq i} \texttt{b}^{\textsf{K}, \textsf{K}_1, \textsf{K}_2}_{jkl} \nonumber \\
	& & \qquad \times F_4(R_{ij}, R_{ik}, R_{il}, \cdots; \theta_{ijk}, \theta_{ikl}, \cdots).
\end{eqnarray}
It is worth noting that most of the symmetry functions also depend on the IR indices $\textsf{K}$ of the internal symmetry group. We also note that since the relative angles $\theta_{ijk}$ are pre-defined constants for models on a regular lattice, the dependence of the $F$ functions on these angles is trivial. More importantly, these $F$ functions are used to select the more relevant pairs or triplets to be included in the symmetry functions. 

In particular, the symmetry functions can be simplified to a sum over the symmetric-IR for lattice models. Take $G_{3b}$ as an example, we first divide all atomic pairs $(jk)$ in the neighborhood into inequivalent classes such that pairs  within the same class are related by the point group symmetry. Moreover, since pairs belong to the same class are related by rotations or reflections that preserve the distance from the center site, they share the same value of the $F_3$ function; see Fig.~\ref{fig:blocks}(a) for an example of the symmetry-related pairs on a square lattice. Using $\pi$ to denote the inequivalent classes of pairs, we then have
\begin{eqnarray}
	G^{\textsf{K}_1, \textsf{K}_2}_{3b}(\{\xi_m\}) = \sum_{\pi} F'_3(\pi; \{\xi_m\}) \sum_{\hat{g}} \texttt{p}^{\textsf{K}_1, \textsf{K}_2}_{\pi(\hat{g})}.
\end{eqnarray}
Here $\pi(\hat{g})$ denotes atomic pairs $(jk)$ related to a reference pair in the class $\pi$ by the symmetry operation $\hat{g}$.
The sum over $\hat{g}$, which is the symmetric sum of the pair-wise invariants $\texttt{p}$, corresponds to the 1D trivial IR of the lattice point group. Consequently, the symmetry function $G_{3b}$ is manifestly an invariant of both the internal and lattice symmetry groups. 

To briefly conclude this Section, we have formulated a general theory of descriptors for characterizing dynamical classical fields in condensed matter systems, and presented various different, yet related, approaches for computing the invariant feature variables. By generalizing the concept of the Weyl matrix for atomic environment, we show that the ordered eigenvalues of a correlation matrix can be used to characterize the classical fields in a local neighborhood. The group-theoretical method offers a rigorous and systematic approach to derive a descriptor based on the bispectrum coefficients. Finally, we discuss a descriptor that incorporates the symmetry of the classical fields into the atom-centered symmetry functions. Explicit implementations of these descriptors are demonstrated for well-studied correlated electron systems in the following sections.

\section{Example: Adiabatic Dynamics of classical scalar field}

\label{sec:scalar}

We first discuss descriptors for the simplest classical field: a dynamical scalar variable $Q_i = Q(\mathbf r_i)$ associated with every lattice sites in a square lattice. Physically, such dynamical scalar field can be viewed as describing the local isotropic structure distortion, for example, the breathing mode of the MO$_6$ octahedron in transition metal oxide. Specific example is given by the Holstein model~\cite{holstein59} with spinless electrons:
\begin{eqnarray}
	\label{eq:H_holstein}
	\hat{\mathcal{H}}=-t \sum_{\langle i j \rangle} \left( \hat{c}_{i}^{\dagger} \hat{c}_{j} + {\rm h.c.} \right) - g\sum_{i} Q_{i} \hat{n}^{\,}_i  
\end{eqnarray}
where $\hat{c}_{i}/\hat{c}^\dagger_i$ is the annihilation/creation operators of spin-less electron at site-$i$, and $\hat{n}_i = \hat{c}_i^\dagger \hat{c}^{\,}_i$ is the corresponding number operator. The first-term describes electron hopping between nearest-neighbor sites $\langle ij \rangle$, $t$ is the nearest-neighbor hopping coefficient. The second term denotes phonon-electron interaction with a coupling constant $g$. The Hamiltonian is supplemented by the classical potential energy that describes the elastic energies of the local structural distortion, 
\begin{eqnarray}
	\mathcal{V}(\{Q_i\}) = \frac{K_0}{2} \sum_{i}  Q_{i}^2 + K_1 \sum_{\langle ij \rangle}  Q_i Q_j. 
\end{eqnarray}
where $K_0$ and $K_1$ are the effective spring constants.
Inclusion of electron-electron interaction leads to the Holstein-Hubbard model~\cite{zhong92} with spinful electrons
\begin{eqnarray}
	\label{eq:H_hols_hub}
	& & \hat{\mathcal{H}} = -t \sum_{\langle ij \rangle}\sum_{\sigma=\uparrow,\downarrow} \left( \hat{c}_{i, \sigma}^{\dagger} \hat{c}_{j, \sigma} + {\rm h.c.} \right)    \\
	& & \quad + U\sum_i \hat{n}_{i, \uparrow} \hat{n}_{i, \downarrow} + V\sum_{\langle ij \rangle} \hat{n}_i \hat{n}_j - g\sum_{i} Q_{i} \hat{n}^{\,}_i, \nonumber
\end{eqnarray}
Here $\hat{n}_{i,\sigma} = \hat{c}^\dagger_{i, \sigma} \hat{c}^{\,}_{i, \sigma}$ is the number operator of electron with spin-$\sigma$, and $\hat{n}_i = \hat{n}_{i,\uparrow} + \hat{n}_{i, \downarrow}$, the $U$ term describes the well known on-site Hubbard repulsion, and $V$ represents short-range Coulomb interactions.  

The Holstein models in which the lattice degrees of freedom are treated quantum mechanically are used to study phenomena related to electron-phonon coupling, such as polaron physics and superconductivity. On the other hand, Holstein models with classical phonons also serve as simple model systems to investigate the effects of structural distortions on the electronic properties. Indeed, the Jahn-Teller model, which is the multi-orbital generalization of the Holstein model, plays an important role in the physics of colossal magnetoresistance effect. In particular, as discussed in Sec.~\ref{sec:intro}, complex inhomogeneous states can arise from the interplay between the fast electron and slow classical lattice dynamics. 

Here we are interested in the adiabatic dynamics of these models and treat the lattice distortions as classical dynamical variables. Their time evolution is then governed by the Langevin equation
\begin{eqnarray}
	\label{eq:langevin}
	\mu \frac{d^2 Q_i}{dt^2} + \lambda \frac{d Q_i}{dt} = -\frac{\partial \mathcal{V}}{\partial Q_i} - \frac{\partial \langle \hat{\mathcal{H}} \rangle}{\partial Q_i} + \eta_i(t).
\end{eqnarray}  
where $\mu$ is the effective mass and $\lambda$ is the dissipation constant, and $\eta_i(t)$ represents the stochastic thermal forces. The first term on the right hand side describes the classical elastic restoring force, while the second term is due to the electron-lattice coupling. Explicit calculation gives
\begin{eqnarray}
	F^{\rm elec}_i = -\frac{\partial \langle \hat{\mathcal{H}} \rangle}{\partial Q_i} = g \langle \hat{n}_i \rangle,
\end{eqnarray}
The electron force is proportional to the on-site electron density. Similar to the Born-Oppenheimer approximation in {\em ab initio} or quantum MD simulations, the electrons are assumed to quickly reach quasi-equilibrium of the instantaneous Hamiltonian. For a given classical field, the Holstein model describes a quadratic fermionic Hamiltonian, which can be solved by, e.g. exact diagonalization. In the presence of Hubbard interaction $U$, many-body methods such as the real-space Gutzwiller/slave boson~\cite{ma19}, or DMFT are required to solve the electron Hamiltonian and compute the electron force. As discussed in Sec.~\ref{sec:framework}, these are time-consuming computations for large system sizes, and ML methods can be employed to achieve large-scale dynamical simulations. In the following, we implement the descriptors discussed in Sec.~\ref{sec:descriptor} to the Holstein-type models. 

\subsection{Correlation matrix}

We first discuss the descriptor based on the correlation matrix. Since here we are dealing with a scalar field, there is no internal index.  Using the notations introduced to label the neighborhood sites in Sec.~\ref{sec:descriptor}, we arrange the lattice distortions into a vector $\mathcal{U}$ with elements $\mathcal{U}_{\textsf{r}} = Q_{j_{\textsf{r}}} = Q(\mathbf r_{j_\textsf{r}})$. The matrix index $\textsf{r} = 1, 2, 3, \cdots, L$, where $L$ is the total number of sites in the neighborhood. We reserve $\textsf{r} = 1$ for the center site, i.e. $j_1 = i$. For convenience, we also define $ R_{\textsf{r} \textsf{s}} = | \mathbf r_{j_\textsf{r}} - \mathbf r_{j_\textsf{s}} |$. The explicit definition of the correlation matrix is (c.f. Eq.~(\ref{eq:corr-m})) 
\begin{eqnarray}
	\begin{array}{ll}
	C_{11} =  \mathcal{U}_1^2 \\ 
	C_{\textsf{r}\textsf{r}} =     \mathcal{U}_\textsf{r}^2 / R_{1\textsf{r}}^2 \quad &  (\textsf{r} \neq 1) \\ 
	C_{\textsf{r}1} = C_{1\textsf{r}} = \mathcal{U}_1 \, \mathcal{U}_\textsf{r} /R_{1\textsf{r}}^2 \quad & (\textsf{r} \neq 1) \\
	C_{\textsf{r}\textsf{s}} = C_{\textsf{s}\textsf{r}} = \mathcal{U}_\textsf{r}\,\mathcal{U}_\textsf{s} / R_{1\textsf{r}} R_{1\textsf{s}} R_{\textsf{r}\textsf{s}} \quad & (\textsf{r},\textsf{s} \neq 1, \textsf{r} \neq \textsf{s}) 
	\end{array}
\end{eqnarray}
The dimension of the correlation matrix is given by the number of lattice sites $L$ in the local neighborhood, which in our implementation contains a total of $L = 89$ sites (up to the 14th neighbors). Since this is a relatively small number, all eigenvalues of the $C$ matrix are used for the descriptor.

\begin{figure}
\includegraphics[width=1.0\columnwidth]{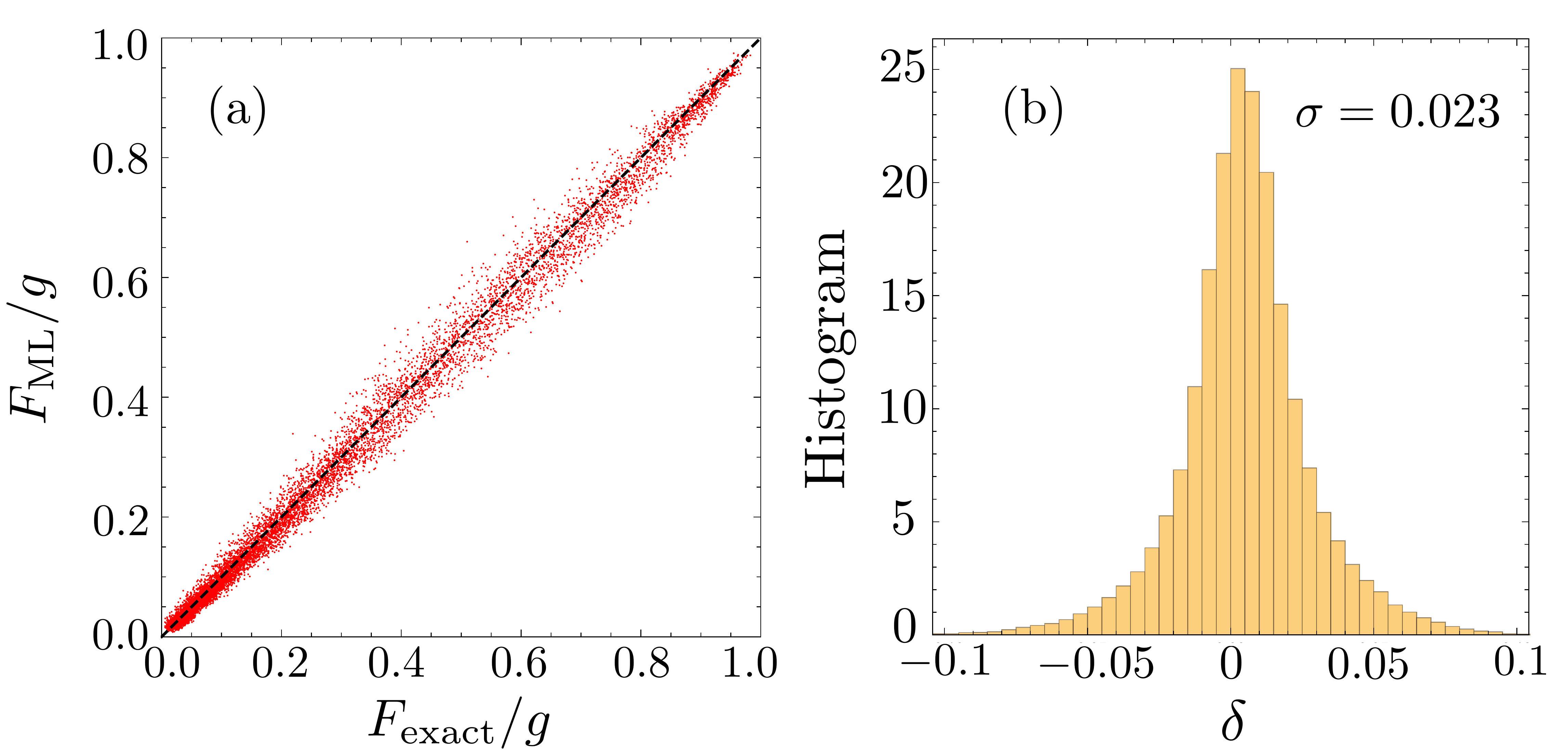}
\caption{(a) Electron forces predicted from ML model with the correlation-matrix descriptor versus exact solutions for test dataset of the Holstein model with $g = 1.5 t$. Here the forces are normalized by the coupling constant $g$, hence are the same as the on-site electron density $n_i$. (b) Histogram of the force error $\delta = (F_{\rm ML} - F_{\rm exact})/g$.}
\label{fig:c-mat}  \
\end{figure}

We integrate the correlation-matrix descriptor with a neural network (NN) learning model to predict the electron force. As a proof of principle, we consider the Holstein model Eq.~(\ref{eq:H_holstein}) without electron-electron interaction. By exactly diagonalizing the quadratic Hamiltonian, the electron force, which is proportional to the on-site electron density, is obtained from the eigenvectors. A six-layer NN model is trained from 2000 snapshots of a $30\times 30$ system. Fig.~\ref{fig:c-mat} shows the ML predictions versus the exact forces. Here we plot the dimensionless forces normalized by the coupling constant $g$, which is the same as the on-site electron density.  The histogram of the prediction error exhibits a small standard deviation $\sigma = 0.023$, indicating very good accuracy of the ML predicted forces.

A remark about the ML model. The results shown in Fig.~\ref{fig:c-mat} were obtained from an ML energy model based on the BP scheme shown in Fig.~\ref{fig:ml-potential}. The output of the NN is the local energy $\epsilon_i$, and the force is obtained via automatic differentiation of the total energy. Since the scalar force is simply proportional to the electron density, one can apply the supervised learning to build a NN which directly predicts the on-stie density $\langle \hat{n}_i \rangle$ from the lattice distortions $\{Q_j\}$ in the neighborhood.  Interestingly, we found that the accuracy of this direct approach is worse than that based on the BP method. As already noted in previous works, the BP method ensures that the predicted forces are conservative as they are given by the derivative of an effective energy. The more constrained supervised  learning in the BP scheme also helps with the prediction accuracy.

\subsection{Bispectrum}

\label{sec:Q-bispectrum}

\begin{table}[b]
\bigskip
\begin{tabular}{|c|c|c|c|c|c|c|c|}
\hline
$D_{4}$ & $E$ & $2C_{4}(z)$ & $C_{2}(z)$ & $2C^{'}_{2}$ & $2C^{''}_{2}$ & Linear  func & Quadaratic  func  \\ \hline \hline
$A_{1}$ & +1  & +1          & +1         & +1           & +1  & -- & $(x^2+y^2)$, $z^2$           \\ \hline
$A_{2}$ & +1  & +1          & +1         & $-1$           & $-1$  & $z$ & --           \\ \hline
$B_{1}$ & +1  & $-1$          & +1         & +1           & $-1$  & -- & $x^2 - y^2$          \\ \hline
$B_{2}$ & +1  & $-1$          & +1         & $-1$           & +1  & -- & $xy$           \\ \hline
$E$     & +2  & 0           & $-2$         & 0            & 0  & $(x, y)$ & $(xz, yz)$            \\ \hline
\end{tabular}
\caption{\label{table:D4-charac} Character table for point group D$_{4}$. Also shown are the linear and quadratic function representations of the various IRs.}
\end{table}

\begin{figure}[t]
\includegraphics[width=0.95\columnwidth]{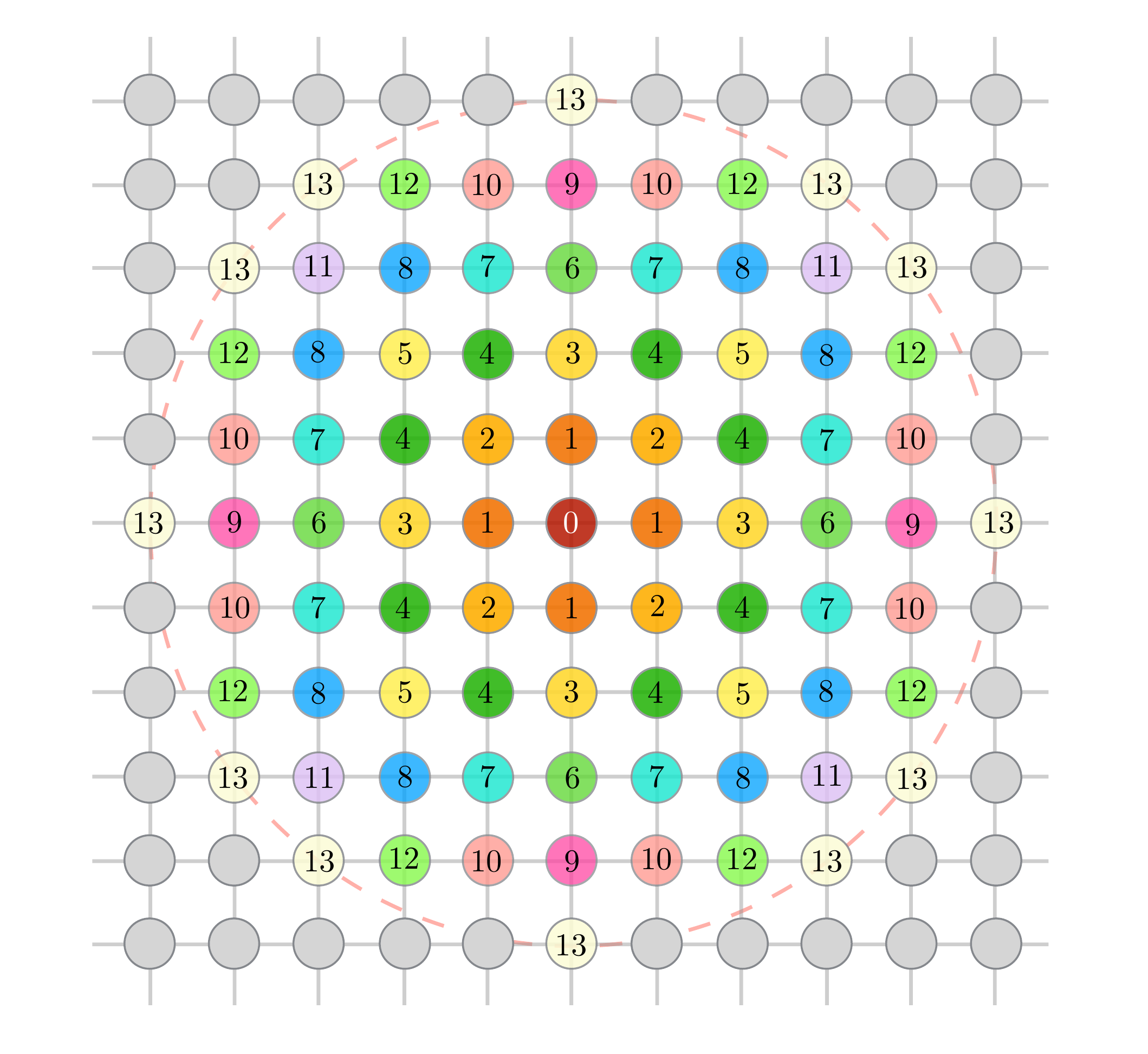}
\caption{Partition of lattice sites in the neighborhood $\mathcal{C}_i$ into groups (nearest neighbors, 2nd nearest neighbors, and so on) depending on their distance to the center site. Lattice sites belong to the same neighboring group are also related by symmetry operations of the point group.  }
\label{fig:lattice}  \
\end{figure}

Next we discuss the bispectrum descriptor of the Holstein-type models based on the group-theoretical method.   The site-symmetry of the square lattice is described by the D$_4$ point group. As discussed above, the collection of on-site lattice distortions $\{Q_j\}$ in the neighborhood  forms a high-dimensional representation of the D$_4$ group, which can be decomposed into the five irreducible representations: $A_1$, $A_2$, $B_1$, $B_2$, and $E$; see Table~\ref{table:D4-charac} for the character table of the point group D$_4$. The first four are singlet IRs, while $E$ is a doublet representation. The task of the decomposition is made easier by noting that the $\{Q_j\}$ of same radius from the center form invariant blocks under the symmetry operations, i.e. the matrix representation of symmetry operations of D$_4$ are block-diagonalized with each block corresponding to a given radius; see Fig.~\ref{fig:lattice}.

\begin{figure}
\includegraphics[width=0.95\columnwidth]{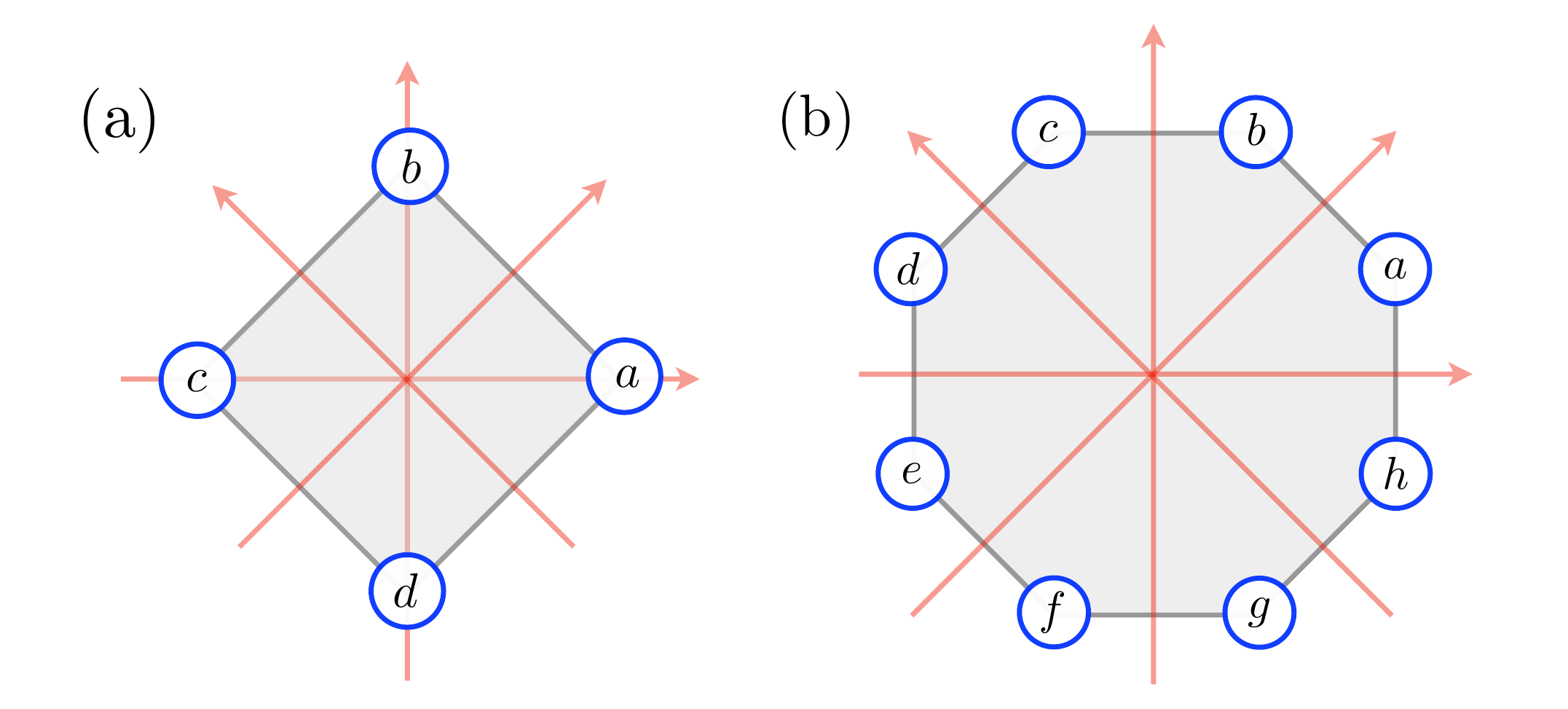}
\caption{The two basic types of neighboring groups with (a)~4 and (b)~8 sites. These groups form invariant blocks in the representations of the neighborhood classical field.  }
\label{fig:inv-block}  \
\end{figure}

In fact, direction examination shows that there are only two kinds of invariant blocks, one of size 4 and the other 8, illustrated in Fig.~\ref{fig:inv-block}(a) and (b), respectively. Consequently, one only needs to decompose the resultant 4- and 8-dimensional reducible representations. The decomposition of the 4-site blocks in Fig.~\ref{fig:inv-block}(a) is $4  =  A_1 \oplus  B_1 \oplus  E$, with the following coefficients:
\begin{eqnarray}
	\label{eq:decomp-4site}
	& &  f^{A_1} = Q_a + Q_b + Q_c + Q_d, \nonumber \\
	& & f^{B_1} = Q_a - Q_b + Q_c - Q_d,  \\
	&  & f^{E}_1 = Q_a - Q_c, \quad f^{E}_2 = Q_b - Q_d. \nonumber
\end{eqnarray}
The decomposition of the 8-site block shown in Fig.~\ref{fig:inv-block}(b) is: $8 = A_1 \oplus A_2 \oplus B_1 \oplus  B_2 \oplus 2 E$. Importantly, there are two doublet $E$ IRs. The corresponding coefficients are
\begin{eqnarray}
	\label{eq:decomp-8site}
	& &  f^{A_1} = Q_a + Q_b + Q_c + Q_d + Q_e + Q_f + Q_g + Q_h, \nonumber  \\
	& &  f^{A_2} = Q_a - Q_b + Q_c - Q_d + Q_e - Q_f + Q_g - Q_h,  \nonumber \\
	& &  f^{B_1} = Q_a - Q_b - Q_c + Q_d + Q_e - Q_f - Q_g + Q_h,  \nonumber \\
	& & f^{B_2} = Q_a + Q_b - Q_c - Q_d + Q_e + Q_f - Q_g - Q_h,\nonumber  \\
	& & f^{E}_1 = Q_a - Q_e, \qquad f^{E}_2 = Q_c - Q_g \nonumber \\
	& & f^{E'}_1 = Q_b - Q_f, \qquad f^{E'}_2 = Q_d - Q_h.
\end{eqnarray}
 Since for most point groups, the dimension of the IRs is often very small, and IR of the same transformation properties appears many times in the decomposition of the vector $\vec{\mathcal{U}}$ representation of the neighborhood, we label the IR index as $\Gamma = (\mathbb{T}, r)$, where $\mathbb{T}$ denotes the symmetry type of the IR, and $r$ enumerates the multiple occurrence of this symmetry in the decomposition. 
Using the above formulas for the different neighborhood blocks, we thus decompose the $\{Q_j\}$ variables into five different IRs $f^{(A_1, r)}$, $f^{(A_2, r)}$, $f^{(B_1, r)}$, $f^{(B_2, r)}$, and $\bm f^{(E, r)} = (f^{(E, r)}_1, f^{(E, r)}_2)$.

Now that we have all the IR components, we can now use Eq.~(\ref{eq:bispectrum1}) to compute the bispectrum coefficients. To this end, we first list in Table~\ref{table:D4-product} the decomposition of tensor product for the $D_4$ group. The nonzero Clebsch-Gordan (CG) coefficients of the tensor products can be found in, e.g. Ref.~\cite{rykhlinskaya06}. The bispectrum coefficients can be classified according to the tensor-product table. First, we list the coefficients involving only the singlets:
\begin{subequations}\label{eq:b_singlet}
\begin{eqnarray}
	\label{eq:b_A1A1A1}
	& & b^{(A_1,  A_1, A_1)}_{r, r', r''} = f^{(A_1, r)} f^{(A_1', r')} f^{(A_1'', r'')},  \\
	\label{eq:b_A1A2A2}
	& & b^{(A_1, A_2, A_2)}_{r, r', r''} = f^{(A_1, r)} f^{(A_2, r')} f^{(A_2, r'')}, \\
	\label{eq:b_A1B1B1}
	& & b^{(A_1, B_1, B_1)}_{r, r', r''} = f^{(A_1, r)} f^{(B_1, r')} f^{(B_1, r'')}, \\
	\label{eq:b_A1B2B2}
	& & b^{(A_1, B_2, B_2)}_{r, r', r''} = f^{(A_1, r)} f^{(B_2, r')} f^{(B_2, r'')}, \\
	\label{eq:b_A2B1B2}
	& & b^{(A_2, B_1, B_2)}_{r, \, r', \, r''} =  f^{(A_2, r)} f^{(B_1, r')} f^{(B_2, r'')}.
\end{eqnarray}
\end{subequations}
There are four different types of bispectrum coefficients involving the doublet. Their expressions can be simplified using the Pauli matrices ${\bm \sigma}_{1, 2, 3}$:
\begin{subequations} \label{eq:b_doublet}
\begin{eqnarray}
	\label{eq:b_A1EE}
	& & b^{(A_1, E, E)}_{r, r', r''} = f^{(A_1, r)} \Bigl( f^{(E, r')}_1 f^{(E, r'')}_1 + f^{(E, r')}_2 f^{(E, r'')}_2 \Bigr) \nonumber \\
	& & \qquad \quad \quad = f^{(A_1, r)} \, {\bm f}^{(E, r')} \cdot {\bm f}^{(E, r'')},  \\ 
	\label{eq:b_A2EE}
	& & b^{(A_2, E, E)}_{r, r', r''} = f^{(A_2, r)} \Bigl( f^{(E, r')}_1 f^{(E, r'')}_2 - f^{(E, r')}_2 f^{(E, r'')}_1 \Bigr) \nonumber \\
	& & \qquad  \quad \quad =  f^{(A_2, r)} \,  {\bm f}^{(E, r')} \cdot (-i  {\bm \sigma}_2) \cdot {\bm f}^{(E, r'')}   ,  \\
	\label{eq:b_B1EE}
	& & b^{(B_1, E, E)}_{r, r', r''} = f^{(B_1, r)} \Bigl( f^{(E, r')}_1 f^{(E, r'')}_1 - f^{(E, r')}_2 f^{(E, r'')}_2 \Bigr) \nonumber \\
	& & \qquad \quad \quad =  f^{(B_1, r)} \,  {\bm f}^{(E, r')} \cdot {\bm \sigma}_3 \cdot {\bm f}^{(E, r'')}  ,  \\
	\label{eq:b_B2EE}
	& & b^{(B_2, E, E)}_{r, r', r''} = f^{(B_2, r)} \Bigl( f^{(E, r')}_1 f^{(E, r'')}_2 + f^{(E, r')}_2 f^{(E, r'')}_1 \Bigr) \nonumber \\
	& & \qquad \quad \quad =  f^{(B_2, r)} \,  {\bm f}^{(E, r')} \cdot {\bm \sigma}_1 \cdot {\bm f}^{(E, r'')} ,
\end{eqnarray}
\end{subequations}
While bispectrum provides a complete description of the neighborhood within a cutoff, a formal descriptor based on bispectrum requires a large number of $b$ coefficients, which makes it infeasible practically. Besides, the various coefficients $b$ are not independent of each other.  To simplify the calculation, our approach here is to use the power spectrum $p^{\Gamma}_r$ supplemented by some of the bispectrum coefficients to obtain an equivalent, but more efficient, descriptor.

\begin{table}[t]
\begin{ruledtabular}
 \begin{tabular}{c c c c c c c} 
  & \quad & $A_1$ & $A_2$ & $B_1$ & $B_2$ & $E$ \\
 \hline
 $A_1$ & \quad & $A_1$ & $A_2$ & $B_1$ & $B_2$ & $E$ \\
 $A_2$ & \quad & & $A_1$ & $B_2$ & $B_1$ & $E$ \\
 $B_1$ & \quad & & & $A_1$ & $A_2$ & $E$ \\
 $B_2$ & \quad & & & & $A_1$ & $E$ \\
 $E$ & \quad & & & & & $A_1 \oplus A_2 \oplus B_1 \oplus B_2$ \\
\end{tabular}
\end{ruledtabular}
\caption{\label{table:D4-product} Direct products of irreducible representations of the D$_4$ point group.} 
\end{table}

\subsection{Reference coefficient for irreducible representations}
\label{sec:ref-IR}

The bispectrum provides a systematic method to obtain invariants which contain crucial information regarding the relative ``phases" between different IRs. However, the set of all bispectrum coefficients listed in Eqs.~(\ref{eq:b_singlet}) is obviously over-complete. For example, since $f^{(A_1)}_r$ is already an invariant itself, the $b^{(A_1, A_1, A_1)}_{r, r', r''}$ is redundant. Here we propose a novel method to retain the phase information based on the idea of {\em reference} coefficients for irreducible representations of the site-symmetry group. Importantly, this method allows us to significantly reduce the number of feature variables required to reconstruct the environment configuration module the site symmetry. 

To demonstrate the idea of reference coefficients, we consider bispectrum $b^{(A_1, \Gamma, \Gamma)}$, where $\Gamma = A_2, B_1$, or $ B_2$ is one of the singlet IR. For convenience, we define the phase of the singlet coefficient as 
\begin{eqnarray}
	f^{(\Gamma, r)} = \sqrt{p^{\Gamma}_r} \, \eta^{\Gamma}_r,  
\end{eqnarray}
Since the $A_1$ part is already invariant under symmetry operations, the invariance of $b^{(A_1, \Gamma, \Gamma)}$ is equivalent to the invariance of the following product
\begin{eqnarray}
	 f^{(\Gamma, r)} \, f^{(\Gamma, r')} = \sqrt{  p^{\Gamma}_{r} p^{\Gamma}_{r'}} \,\eta^{\Gamma}_{r, r'}
\end{eqnarray}
where we have defined the relative phase of two expansion coefficients as
\begin{eqnarray}
	\eta^{\Gamma}_{r_1, r_2} = \eta^{\Gamma}_{r_1} \, \eta^{\Gamma}_{r_2}.
\end{eqnarray}
In addition to the power spectrum coefficients $p^{\Gamma}$, the relative phase $\eta^{\Gamma}_{r_1, r_2}$ is a crucial invariant encoded in the bispectrum.
However, the relative phases are not independent of each other. Indeed, from its definition it is straightforward to show that 
\begin{eqnarray}
	\label{eq:relative-phase}
	\eta^{\Gamma}_{r_1, r_2} \, \eta^{\Gamma}_{r_2, r_3} \, \eta^{\Gamma}_{r_3, r_1} = 1.
\end{eqnarray}
To derive the set of truly independent phase coefficients, we introduce the phase $\eta^{\Gamma}_*$ of a reference expansion coefficient $f^{(\Gamma, *)}$ of IR-$\Gamma$, and define the relative phase between a given coefficient $f^{(\Gamma, r)}$ and the reference one as $\eta^{\Gamma}_{*, r}$.  The relative phase between two expansion coefficients of the same IR $\Gamma$ is then given by
\begin{eqnarray}
	\label{eq:relative-phase2}
	\eta^{\Gamma}_{r_1, \, r_2} = \eta^{\Gamma}_{*, \, r_1} \, \eta^{\Gamma}_{*, \, r_2}.
\end{eqnarray}
This relation thus allow us to reconstruct all the $b^{(A_1, \Gamma, \Gamma)}$ bispectrum coefficients.

\begin{figure}[t]
\includegraphics[width=1.0\columnwidth]{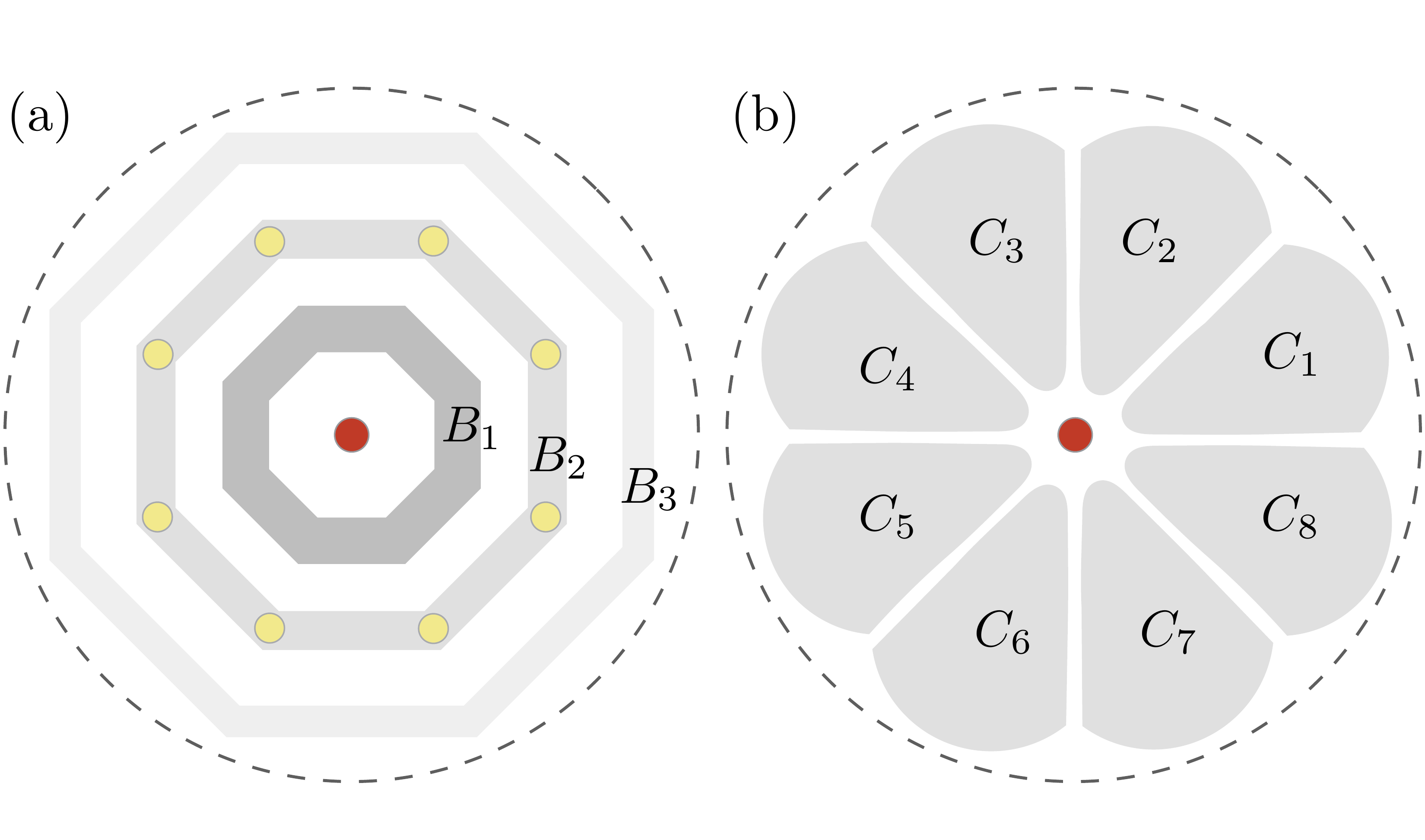}
\caption{Construction of the reference irreducible representations from (a) a particular neighboring group, and (b)~symmetry-related clusters.}
\label{fig:ref-irrep}  \
\end{figure}

It is worth noting that the purpose of the reference coefficient of a given IR is to construct invariants which retain the relative phase between a given expansion coefficient and the reference one. Consequently, the amplitude of the reference representations is completely irrelevant, as long as it is nonzero. Also importantly, there is no unique procedure to obtain these reference expansion coefficients $f^{(\Gamma, *)}$. 
Fig.~\ref{fig:ref-irrep} shows two examples of computing the reference coefficients from the neighborhood configuration. One approach is simply to use the decomposition of a particular 8-site neighbor block, e.g. the $B_2$ neighbors in Fig.~\ref{fig:ref-irrep}(a), as the reference. 
However, the small size of such an 8-site block might result in the undesirable situation of vanishing expansion coefficients for some IRs. While there is no general method to avoid such pathological situation, one can try to minimize such probability, which is already very small in most cases, by using weighted results from several neighbor blocks to derive the reference expansion coefficients. 

Another approach, shown in Fig.~\ref{fig:ref-irrep}(b), is based on symmetry-related clusters built from the neighborhood sites. First, a cluster of sites, say $C_1$, is introduced either based on geometrical consideration, or simply randomly. Applying symmetry operations of the point group to $C_1$ then generates all clusters $C_K$ that are related to $C_1$ by the site symmetry. Next we define the average of dynamical variables for each cluster:  $ \mathcal{Q}_K  = (1/M)\sum_{i \in C_K} Q_i$, where $M$ is the number of sites in each cluster. Obviously, these cluster-based variables $\{\mathcal{Q}_K\}$ form an 8-dimensional reducible representation. Its decomposition using Eq.~(\ref{eq:decomp-8site}) thus gives rise to a set of references coefficients $f^{(\Gamma, *)}$ for all five IRs of the D$_4$ group.

As discussed above, the amplitude of the reference expansion coefficients is unimportant for the descriptor, as the essential information is carried by their phase.  In the following, we use the phase of the reference coefficients to define a new set of feature variables $\{g^{(\Gamma, r)}_r \}$ which are manifestly invariant under symmetry operations. Importantly, these coefficients supplemented by a few additional normalized bispectrum factor form a complete descriptor such that the local environment can be faithfully reconstructed from them. First, since expansion coefficients of the $A_1$ IR is already an invariant, we define $g^{(A_1, r)} = f^{(A_1, r)}$. For other singlet IRs $\Gamma = A_2, B_1$, and $B_2$, the new coefficient is defined as
\begin{eqnarray}
	g^{(\Gamma, r)} = f^{(\Gamma, r)} \, \eta^{\Gamma}_* = \sqrt{p^{\Gamma}_r} \, \eta^{\Gamma}_{*, r}
\end{eqnarray}
where $\eta^{\Gamma}_{*, r}$ is the relative phase introduced in Eq.~(\ref{eq:relative-phase2}). This coefficient is obviously an invariant of the symmetry group. Importantly, the bispectrum coefficients in Eq.~(\ref{eq:b_A1A1A1})--(\ref{eq:b_A1B2B2}) can be readily expressed in terms of these invariant $g$ coefficients:
\begin{eqnarray}
	\label{eq:b_singlet2}
	b^{(A_1, \Gamma, \Gamma)}_{r, r', r''} = g^{(A_1, r)} g^{(\Gamma, r')} g^{(\Gamma, r'')}, 
\end{eqnarray}
where $\Gamma = A_1, A_2, B_1, B_2$. We still need to consider the special bispectrum $b^{(A_2, B_1, B_2)}$ in Eq.~(\ref{eq:b_A2B1B2}), which encodes the relative phases between expansion coefficients of the three 1D IRs. However, it is easy to show that expression in Eq.~(\ref{eq:b_A2B1B2}) can be expressed as
\begin{eqnarray}
	\label{eq:b_singlet3}
	b^{(A_2, B_1, B_2)}_{r, r', r''} = g^{(A_2, r)}  g^{(B_1, r')} g^{(B_2, r'')} b^{(A_2, B_1, B_2)}_* ,
\end{eqnarray}
where we have introduced a normalized reference bispectrum
\begin{eqnarray}
	b^{(A_2, B_1, B_2)}_* = \eta^{(A_2)}_* \, \eta^{(B_1)}_* \, \eta^{(B_2)}_*,
\end{eqnarray}
It can be readily checked that this normalized bispectrum of the reference coefficient is invariant under symmetry operation. Eqs.~(\ref{eq:b_singlet2}) and (\ref{eq:b_singlet3}) indicate that all bispectrum coefficients of Eq.~(\ref{eq:b_singlet}) can be restored using invariant coefficients $g^{(\Gamma, r)}$ and $b^{(A_2, B_1, B_2)}_*$.


Next we consider the bispectrum coefficients in Eq.~(\ref{eq:b_doublet}) that involve the doublet $E$ IRs. We define a similar normalized doublet vector $\bm \epsilon_* = (\epsilon^x_*, \epsilon^y_*)$ from the reference coefficients of the doublet IR:
\begin{eqnarray}
	\epsilon^x_* = f^{(E, *)}_1 / \bigl|f^{(E, *)}_1 \bigr|, \qquad \epsilon^y_* = f^{(E, *)}_2 / \bigl| f^{(E, *)}_2 \bigr|.
\end{eqnarray}
The $x$ and $y$ components have amplitude $|\epsilon^{x, y}_*| = 1$. It can be readily checked that this 2-component vector $\bm \epsilon_*$ transforms as a doublet IR under the D$_4$ group. Consequently, $\bm\epsilon_*$ can be used to build invariant coefficients using the the bispectrum formula in Eq.~(\ref{eq:b_doublet}). Specifically, for any given doublet vector $\bm f^{(E, r)}$, we introduce two invariant coefficients $\bm g^{(E, r)} = (g^{(E, r)}_1, g^{(E, r)}_2)$ defined as
\begin{subequations} \label{eq:g_doublet}
\begin{eqnarray}
	g^{(E, r)}_1 &=& \bm \epsilon_* \cdot \bm f^{(E, r)} / \sqrt{2} \nonumber \\
	&=& \left(\epsilon^x_* f^{(E, r)}_1 + \epsilon^y_* f^{(E, r)}_2 \right)/\sqrt{2},  \\
	g^{(E, r)}_2 &=&  \eta^{(B_1)}_* \, \bm \epsilon_* \cdot \bm\sigma_3\cdot \bm f^{(E, r)} / \sqrt{2} \nonumber \\
	&=& \eta^{(B_1)}_* \left( \epsilon^x_* f^{(E, r)}_1 - \epsilon^y_* f^{(E, r)}_2 \right) / \sqrt{2} . 
\end{eqnarray}
\end{subequations}
This can be readily inverted to give
\begin{subequations}
\begin{eqnarray}
	f^{(E, r)}_1 &=& \epsilon^x_* \left(g^{(E, r)}_1 + \eta^{(B_1)}_* g^{(E, r)}_2 \right) / \sqrt{2},  \\
	f^{(E, r)}_2 &=& \epsilon^y_* \left(g^{(E, r)}_1 - \eta^{(B_1)}_* g^{(E, r)}_2 \right) / \sqrt{2}.
\end{eqnarray}
\end{subequations}
By substituting the above expressions for $f^{(E, r)}_{1, 2}$ into Eqs.~(\ref{eq:b_doublet}), we can express the bispectrum coefficients in terms of the invariant $g$ coefficients
\begin{subequations}
\begin{eqnarray}
	b^{(A_1, E, E)}_{r, r', r''} &=& g^{(A_1, r)} \, \bm g^{(E, r')} \cdot \bm g^{(E, r'')}, \\
	b^{(A_2, E, E)}_{r, r', r''} &=& b^{(A_2, B_1, B_2)}_* \, b^{(B_2, E, E)}_* \nonumber \\
	  & &\, \times  g^{(A_2, r)} \, \bm g^{(E, r')} \cdot (-i \bm\sigma_2) \cdot \bm g^{(E, r'')},  \\
	b^{(B_1, E, E)}_{r, r', r''} &= & g^{(B_1, r)} \, \bm g^{(E, r')} \cdot \bm\sigma_1 \cdot \bm g^{(E, r'')}, \\
	b^{(B_2, E, E)}_{r, r', r''} & = & b^{(B_2, E, E)}_* g^{(B_2, r)} \, \bm g^{(E, r')} \cdot \bm\sigma_3 \cdot \bm g^{(E, r'')}. \quad
\end{eqnarray}
\end{subequations}
Here we have defined another important normalized reference bispectrum coefficient
\begin{eqnarray}
	b^{(B_2, E, E)}_* = \eta^{(B_2)}_* \, \epsilon^x_* \epsilon^y_*.
\end{eqnarray}
which encodes the relative phase between the reference $B_2$ and $E$ coefficients.

\begin{figure}[t]
\includegraphics[width=1.0\columnwidth]{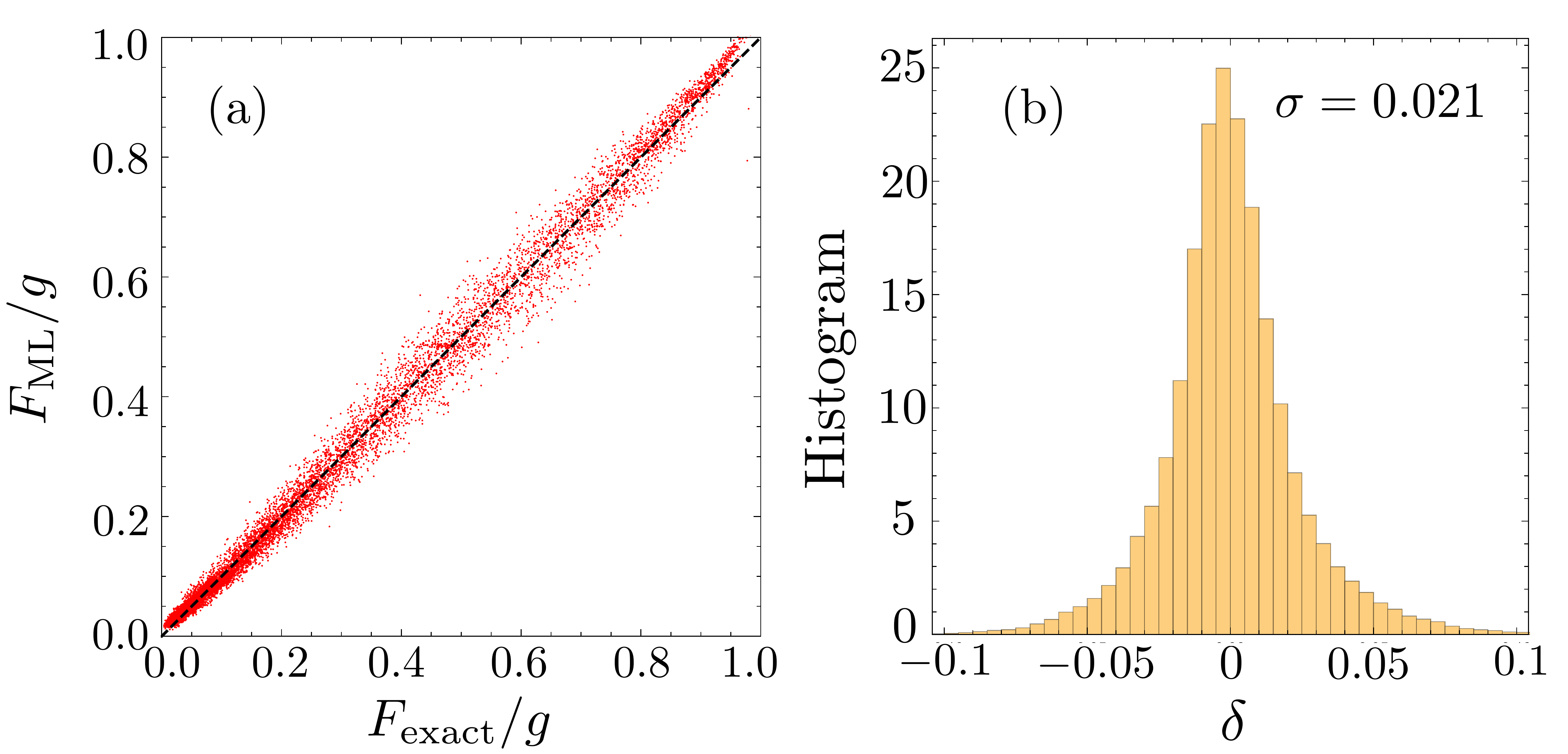}
\caption{(a) Electron forces predicted from ML model with the bispectrum descriptor versus exact solutions for test dataset of the Holstein model with $g = 1.5 t$. Here the forces are normalized by the coupling constant $g$, hence are the same as the on-site electron density $n_i$. (b) Histogram of the force errors $\delta = (F_{\rm ML} - F_{\rm exact})/g$.}
\label{fig:bispec-hols}  \
\end{figure}

To summarize, a complete description of the local environment is given by the following set of invariant coefficients:
\begin{eqnarray}
	& & \Bigl\{ g^{(A_1, r)}, \, g^{(A_2, r)}, \, g^{(B_1, r)}, \, g^{(B_2, r)}, \, \bm g^{(E, r)}, \nonumber\\
	& & \qquad \quad b^{(A_2, B_1, B_2)}_*, \, b^{(B_2, E, E)}_* \Bigr\}.
\end{eqnarray}
We integrate this descriptor with a six-layer NN to develop a ML energy model following the framework shown in Fig.~\ref{fig:ml-potential}. The NN is trained from exact solution of the Holstein model~Eq.~(\ref{eq:H_holstein}) on a $30\times30$ square lattice. The ML predicted normalized forces, shown in Fig.~\ref{fig:bispec-hols}, agree very well with the exact values, as evidenced by the rather small standard deviation from the histogram of prediction errors.

It is worth noting that, thanks to the regularity of the lattice geometry and the simplicity of the scalar classical field, decent force prediction can be obtained using a ML model even without a descriptor, namely directly using an array of the scalar variables $\{ Q_j \}$ in the neighborhood as the input. However, the site symmetry is an approximate symmetry in such naive approaches, though the accuracy of the symmetry could be improved by increasing the size of the training dataset. On the other hand, the employment of the lattice descriptor ensures that the ML force field model preserves the lattice symmetry.

\section{Example: Dynamics of Cooperative Jahn-Teller coupling}
\label{sec:JT}

As a second example of the type-I models, we consider the Jahn-Teller (JT) coupling between the $e_g$ electrons and the distortion modes of local MnO$_6$ octahedron in maganites. The JT-coupling, along with the DE mechanism, are important for the physics of colossal magnetoresistance and polaron liquids~\cite{millis96,maezono03,sen06,popovic00}. Here the local distortion of the tetrahedron is characterized by three modes. The symmetric breathing mode $Q_i^1$ is essentially the same as the Holstein phonons discussed in the previous Section~\ref{sec:scalar}. Here we are interested in the dynamics of the asymmetric normal modes that are described by a $e_g$ doublet $\mathbf Q_i = (Q^x_i, {Q}^z_i)$. Here the $Q^x_i$ component has the symmetry of $(x^2 - y^2)$, and $Q^z_i$ has the symmetry of $(3z^2 - r^2)$.  Here we consider the two-orbital electron Hamiltonian with the JT coupling on a square lattice~\cite{sen06,popovic00}
\begin{eqnarray}
	 & & \hat{\mathcal{H}} = -\sum_{\langle ij \rangle} \sum_{\alpha\beta = a, b} \left( t_{ij}^{\alpha\beta} \hat{c}^\dagger_{i\alpha} \hat{c}^{\,}_{j\beta} + {\rm h.c.} \right)  \\
	& & \quad -  g   \sum_i \left[ Q_i^x \left(\hat{c}_{ia}^\dagger \hat{c}^{\,}_{ib} + \hat{c}_{ib}^\dagger \hat{c}^{\,}_{ia} \right) 
	+ Q_i^z  \left( \hat{c}_{ia}^\dagger \hat{c}^{\,}_{ia} - \hat{c}_{ib}^\dagger \hat{c}^{\,}_{ib} \right) \right], \nonumber
\end{eqnarray}
Here $\hat{c}^\dagger_{i \alpha}/\hat{c}_{i, \alpha}$ are creation/annihilation operators of electron at site-$i$ with an $e_g$ orbital-index $\alpha = a$, $b$, corresponding to orbitals $d_{x^2-y^2}$ and $d_{3z^2 - r^2}$, respectively. The superscript $x$, $z$ indicates that the corresponding $Q$-mode couples to the $x$ and $z$ pseudo-spin of the orbitally degenerate $e_g$ electrons  Because of the orbital degrees of freedom, the electron hopping coefficients are anisotropic: $t^{aa}_{ij} = -\sqrt{3} t^{ab}_{ij} = -\sqrt{3} t^{ba}_{ij} = 3 t^{bb}_{ij} = t$ for $\langle ij \rangle$ along the $x$-direction, and $t^{aa}_{ij} = \sqrt{3} t^{ab}_{ij} = \sqrt{3} t^{ba}_{ij} = 3 t^{bb}_{ij} = t$ for $\langle ij \rangle$ along the $y$-direction. 

The adiabatic dynamics of the JT phonons is described by a similar Langevin equation  
\begin{eqnarray}
	\label{eq:langevin2}
	\mu \frac{d^2 \mathbf Q_i}{dt^2} + \lambda \frac{d \mathbf Q_i}{dt} = -\frac{\partial \mathcal{V}}{\partial \mathbf Q_i} - \frac{\partial \langle \hat{\mathcal{H}} \rangle}{\partial \mathbf Q_i} +\bm \eta_i(t).
\end{eqnarray}  
Here $\mu$ is the effective mass, $\lambda$ is the dissipation coefficient, and $\bm\eta_i(t)$ represent stochastic thermal forces. We have also included the classical elastic energy $\mathcal{V}$ of the JT phonons~\cite{sen06,popovic00}. Again, the time-consuming part, which has to be carried out at every time-step, is the calculation of the electron forces:
\begin{eqnarray}
	 & & \mathbf F^{\rm elec}_i = - \frac{\partial \langle \hat{\mathcal{H}} \rangle}{\partial \mathbf Q_i}  \\
	& & \quad = g \left( \langle \hat{c}_{ia}^\dagger \hat{c}^{\,}_{ib} + \hat{c}_{ib}^\dagger \hat{c}^{\,}_{ia} \rangle,  \,
	\langle \hat{c}_{ia}^\dagger \hat{c}^{\,}_{ia} - \hat{c}_{ib}^\dagger \hat{c}^{\,}_{ib}  \rangle  \right). \nonumber
\end{eqnarray}
The two components of the force are given by the on-site two-point correlation function $\rho_{i\alpha, i\beta} = \langle \hat{c}^\dagger_{i,\beta} \hat{c}^{\,}_{i,\alpha} \rangle$, which can be obtained by exact diagonalization for the JT Hamiltonian above. 
Next we outline the bispectrum descriptors for the JT doubles, which can be combined with a learning model to develop an effective classical energy, as outlined in Fig.~\ref{fig:ml-potential} for the adiabatic dynamics of the JT phonons.

\begin{figure}
\includegraphics[width=1.0\columnwidth]{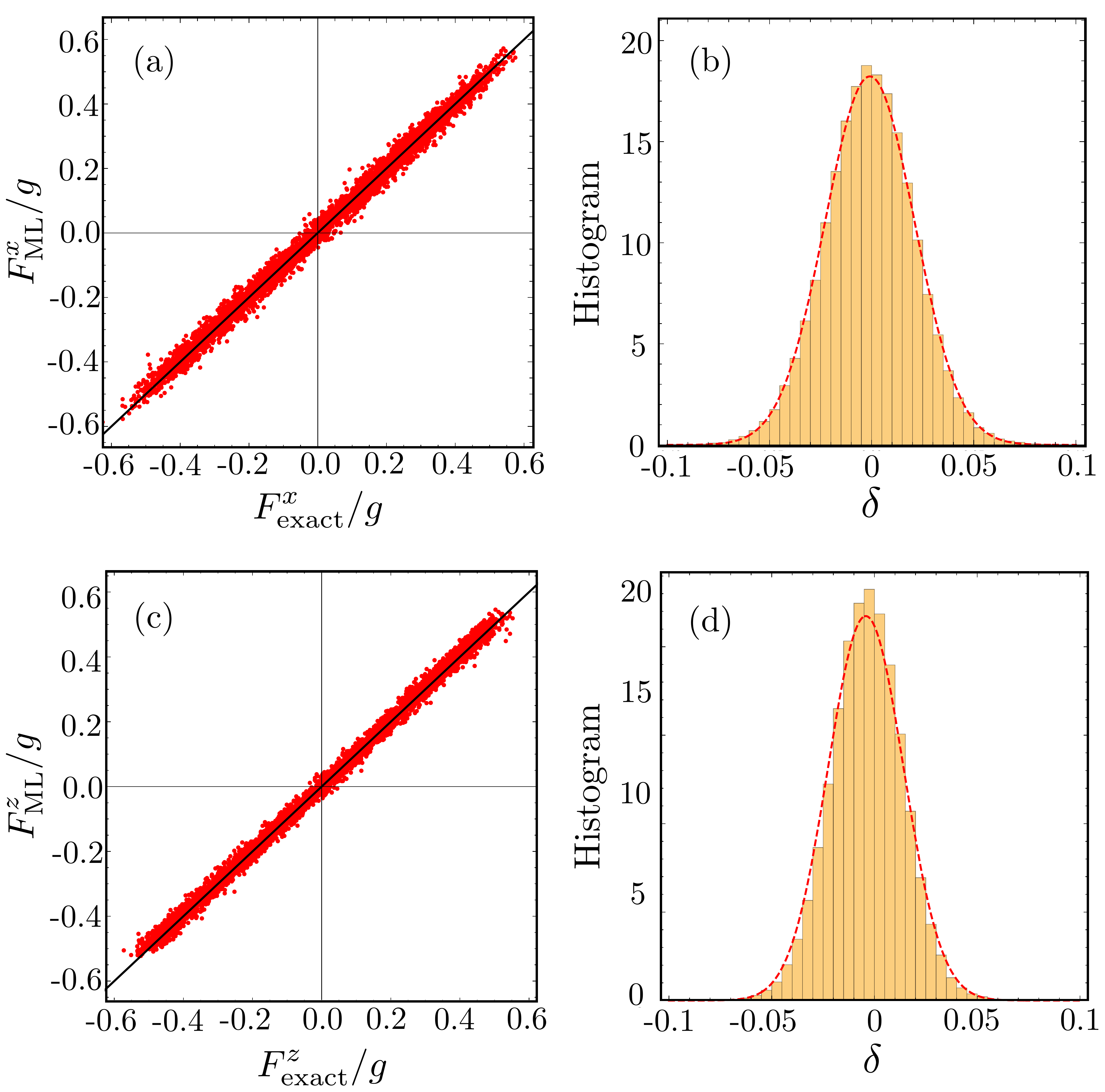}
\caption{Electron forces predicted from ML model with the bispectrum descriptor versus exact solutions for test dataset of the Jahn-Teller model with $g = 1.5 t$ for the (a) $Q^x$ and (c) $Q^z$ components. Here the forces are normalized by the coupling constant $g$, hence are the same as the on-site electron density $n_i$. The corresponding histograms of the force errors $\delta = (F_{\rm ML} - F_{\rm exact})/g$ are shown in panel~(b) and (d), respectively.}
\label{fig:bispec-JT}  \
\end{figure}

First, we note that the JT doublet $\mathbf Q_i$ has the transformation property of an $e_g$ doublet under the cubic point group, e.g. O$_h$. Since here we consider the two-dimensional model, which is relevant for applications of CMR manganites, the two components of $\mathbf Q_i$ transform independently of each other under the in-plane D$_4$ point group of the square lattice. As shown in Table~\ref{table:D4-charac}, while the $Q^z_i$ mode transforms with the $A_1$ symmetry, the $Q^x_i$ mode behaves as the $B_1$ IR. Consequently, the $Q^z_i$ components are essentially a scalar field, and their decomposition can be obtained following exactly the same procedure for the Holstein model outlined in Sec.~\ref{sec:Q-bispectrum}, and the corresponding coefficients are given by Eqs.~(\ref{eq:decomp-4site}) and (\ref{eq:decomp-8site}). On the other hand, the fact that the $Q^x_i$ mode transforms as an $B_1$ IR mixes the transformations of the lattice sites with those of $Q^x$ amplitude. Take the four-site neighboring group in Fig.~\ref{fig:inv-block}(a) as an example, under the $C_4(z)$ rotation, the four $Q^2$ modes acquire an additional $-1$ sign in addition to the lattice rotation:
\begin{eqnarray*}
	\label{eq:xx}
	Q^x_a \to -Q^x_b, \quad Q^x_b \to -Q^x_c \\ Q^x_c \to -Q^x_d, \quad Q^x_d \to -Q^x_a  
\end{eqnarray*}
To account for this additional $-1$ sign, the expansion coefficients of the corresponding IRs are given by
\begin{eqnarray}
	\label{eq:decomp-4site-Qx}
	& & f^{A_1} = Q^x_a - Q^x_b + Q^x_c - Q^x_d,  \nonumber \\
	& &  f^{B_1} = Q^x_a + Q^x_b + Q^x_c + Q^x_d,  \\
	&  & f^{E}_1 = Q^x_a - Q^x_c, \quad f^{E}_2 = Q^x_d - Q^x_b. \nonumber
\end{eqnarray}
Similar expressions can be obtained for the 8-site neighboring block.
\begin{eqnarray}
	\label{eq:decomp-8site-Qx}
		& &  f^{A_1} = Q^x_a - Q^x_b - Q^x_c + Q^x_d + Q^x_e - Q^x_f - Q^x_g + Q^x_h,  \nonumber \\
	& & f^{A_2} = Q^x_a + Q^x_b - Q^x_c - Q^x_d + Q^x_e + Q^x_f - Q^x_g - Q^x_h,\nonumber  \\
	& &  f^{B_1} = Q^x_a + Q^x_b + Q^x_c + Q^x_d + Q^x_e + Q^x_f + Q^x_g + Q^x_h, \nonumber  \\
	& &  f^{B_2} = Q^x_a - Q^x_b + Q^x_c - Q^x_d + Q^x_e - Q^x_f + Q^x_g - Q^x_h,  \nonumber \\
	& & f^{E}_1 = Q^x_a - Q^x_e, \qquad f^{E}_2 = Q^x_g - Q^x_c \nonumber \\
	& & f^{E'}_1 = Q^x_b - Q^x_f, \qquad f^{E'}_2 = Q^x_h - Q^x_d.
\end{eqnarray}
Given the expansion coefficients of both the $Q^x$ and $Q^z$ phonons, bispectrum coefficients, with contributions from both modes, can be obtained as outlined in Sec.~\ref{sec:Q-bispectrum}. The implementation can similarly be simplified using reference IR method. Combining the descriptor with a neural network, an effective energy model is trained by datasets from exact diagonalization of the JT Hamiltonian on $30\times 30$ lattices. The ML predicted forces for both components versus the exact values are shown in Fig.~\ref{fig:bispec-JT} along with the histograms of the prediction errors.

\section{Example: Adiabatic dynamics of itinerant magnets}
\label{sec:spin}

As a final example, in this Section we demonstrate the descriptors for the spin dynamics of itinerant magnets,  which are a representative example of the type-II models. Explicitly, we consider the following single-band s-d model~\cite{yunoki98,zhang04}
\begin{eqnarray}
	\label{eq:H_sd}
	\hat{\mathcal{H}} = -t \sum_{\langle ij \rangle} \left( \hat{c}^{\dagger}_{i \alpha} \hat{c}^{\;}_{j \alpha} + {\rm h.c.} \right)
	- J \sum_{i} \mathbf S_i \cdot \hat{c}^{\dagger}_{i\alpha}  {\bm{\sigma}_{\alpha\beta}} \hat{c}^{\;}_{i\beta}, \qquad
\end{eqnarray}
where $\hat{c}^\dagger_{i \alpha}/\hat{c}_{i, \alpha}$ are creation/annihilation operators of electron with spin $\alpha = \uparrow, \downarrow$ at site $i$,  $ \langle ij \rangle$ indicates the nearest neighbors, $t$ is the electron hopping constant, $J_H$ is the local Hund's rule coupling between electron spin and local magnetic moment $\mathbf S_i$ due to the localized $d$ or $f$ electrons. Here repeated indices $\alpha, \beta$ imply summation. The s-d Hamiltonian offers a fundamental description of the electron-spin interaction; it has been widely used to model the dynamics of magnetic textures, such as domain-walls and skyrmions, under the influence of conducting electrons. The strong coupling $J \gg t$ limit of the s-d Hamiltonian, also known as the double-exchange model, exhibits intriguing phase-separated states in which ferromagnetic clusters with low electron density are mixed with half-filled antiferromagnetic domains~\cite{yunoki98}. Such electronic phase separation is crucial to the emergence of novel material functionalities such as colossal magnetoresistance and high-T$_c$ superconductivity. 

The adiabatic dynamics of the classical spins is governed by the stochastic Landau-Lifshitz-Gilbert (LLG) equation~\cite{brown63,antropov97}
\begin{eqnarray}
	\label{eq:LLG}
	\frac{d \mathbf S_i}{dt}\ = \gamma \mathbf S_i \times \Bigl( \frac{\partial \langle \hat{\mathcal{H}} \rangle}{\partial \mathbf S_i} + \bm\eta_i \Bigr) 
	- \lambda \mathbf S_i \times \Bigl(\mathbf S_i \times \frac{\partial \langle \hat{\mathcal{H}} \rangle}{\partial \mathbf S_i} \Bigr), \qquad
\end{eqnarray}
where $\bm\eta_i(t)$ is a stochastic local field of zero mean, and $\lambda$ is the damping coefficient. The computational overhead is dominated by the calculation of the electron contribution to the exchange forces:
\begin{eqnarray}
	\label{eq:H-force}
	\mathbf H^{\rm elec}_i = - \frac{\partial \langle \hat{\mathcal{H}} \rangle}{\partial \mathbf S_i} = J \bm{\sigma}_{\alpha\beta} \langle \hat{c}^{\dagger}_{i\alpha} \hat{c}^{\;}_{i\beta} \rangle.
\end{eqnarray}
As demonstrated in Fig.~\ref{fig:ml-potential} and in previous works~\cite{zhang20,zhang21}, the calculation of this electronic force can be speed up with the use of ML energy models. Here we present details of the descriptors for the case of dynamical spins, or the vector field. 

The local spins in the standard s-d model are often assumed to have a fixed length. On the other hand, Eq.~(\ref{eq:H_sd}) can also be viewed as an effective Hamiltonian for the spin-density wave (SDW) obtained from a mean-field treatment of the interacting electron Hamiltonian. In the case of the SDW, the spin length itself is a dynamical variable, and the Gilbert damping in Eq.~(\ref{eq:LLG}) has to be replaced by a Langevin-type dissipation force~\cite{chern18}. Here we consider the general case of descriptor for a vector field $\mathbf S(\mathbf r_i)  = (S_i^x, S_i^y, S_i^z)$ with variable spin lengths.

The internal symmetry of the vector field is the SO(3) rotation group, which is locally isomorphic to the SU(2) group. The IRs of the rotation group is labeled by the angular momentum quantum number $j$ which can be either integers or half-integers. The 3-component vector $\mathbf S = (S^x, S^y, S^z)$ is already an IR of dimension 3 of the rotation group, which is equivalent to the angular momentum $j = 1$ representation. In order to construct the bispectrum coefficients of the SO(3) group, we consider the tensor product of two spins $\mathbf S_j \otimes \mathbf S_k$ in the neighborhood. Since this can also be viewed as the tensor product $(j=1) \otimes (j=1)$, from the theory of angular momentum addition, it can be decomposed to a direct sum of $j=0$ (scalar), $j = 1$ (vector), and $j = 2$ (rank-2 traceless symmetric tensor).

The scalar component ($j = 0$) is simply the inner product of the two vectors 
\begin{eqnarray}
	\texttt{p}_{jk} = \mathbf S_j \cdot \mathbf S_k, 
\end{eqnarray}
This scalar invariant is the same as the generalized power spectrum coefficient Eq.~(\ref{eq:p-block}) in Sec.~\ref{sec:descriptor}, which is a special case of the bispectrum coefficients obtained from the trivial representation and two $j=1$ IRs. The invariants $\texttt{p}_{jk}$ with $j \neq k$ represents the rotation-invariant spin-spin correlation, sometimes also called a bond variable.  For the case of SDWs, one also needs to consider the on-site invariant $\texttt{p}_{jj} = |\mathbf S_j|^2$, which corresponds to the spin length.  

The $j = 1$ IR in the decomposition of $\mathbf S_j \otimes \mathbf S_k$ is the vector product $\mathbf S_j \times \mathbf S_k$, which can be combined with a third vector to form a scalar invariant under rotation
\begin{eqnarray}
	\texttt{b}_{jkl} =  \mathbf S_i \times \mathbf S_k \cdot \mathbf S_l.
\end{eqnarray}
These are the bispectrum coefficients Eq.~(\ref{eq:b-block}) constructed from three $j=1$ IRs of the SO(3) group. This scalar invariant $\texttt{b}_{jkl}$, also called the scalar spin chirality, provides a measure of the non-coplanarity of the triplet spins. Importantly, according to the bispectrum theory of group representations, the collection of all scalars $\texttt{p}_{jk}$ and $\texttt{b}_{jkl}$ provides a faithful representation of the neighborhood that is invariant with respect to SO(3) rotations. On the other hand, this is obviously an over-complete representation. For example, the relative orientation between three spins $(ijk)$ can be completely specified by three bond variables $\texttt{p}_{ij}$, $\texttt{p}_{ik}$, and $\texttt{p}_{jk}$; consequently, the scalar chirality $\texttt{b}_{ijk}$ is redundant. This also means that an equally faithful description of three spins can be attained from $\texttt{p}_{ij}$, $\texttt{p}_{ik}$, and $\texttt{b}_{ijk}$. However, the effectiveness of the descriptor depends on the choice of the feature variables, and we find that explicit inclusion of the scalar chirality is very efficient for describing magnetic states with non-coplanar spins. 

\subsection{Atom-centered symmetry functions}

As a first example, we discuss the symmetry functions based on the rotation-invariant building blocks discussed above. Following Eqs.~(\ref{eq:G2a})--(\ref{eq:G3b}) in Sec.~\ref{sec:scsf}, we define the following symmetry functions associated with a center spin at $\mathbf r_i$. First, since there is no on-site singlet (trivial) IR $\texttt{f}_j$ for the spin model, there is no $G_{2a}$. The only nontrivial two-body symmetry function is  
\begin{eqnarray}
	\label{eq:G2b-spin}
	G_{2b}\bigl(\{\xi_m\} \bigr) = \sum_{j \neq i} F'_{2}\bigl( R_{ij}; \{\xi_m\} \bigr) (\mathbf S_i \cdot \mathbf S_j).
\end{eqnarray}
And there are also two 3-body invariants involving a pair of spins $(jk)$ along with the center site-$i$:
\begin{eqnarray}
	\label{eq:G3a-spin}
	& & G_{3a}\bigl(\{\xi_m\} \bigr) = \sum_{jk \neq i} F_{3}\bigl(R_{ij}, R_{ik}, R_{jk}; \theta_{ijk}) (\mathbf S_j \cdot \mathbf S_k), \quad \\
	\label{eq:G3b-spin}
	& & G_{3b}\bigl(\{\xi_m\} \bigr) = \sum_{jk \neq i} F'_{3}\bigl(R_{ij}, R_{ik}, R_{jk}; \theta_{ijk}) (\mathbf S_i \cdot \mathbf S_j \times \mathbf S_k).\, \qquad
\end{eqnarray}
We note that these symmetry functions can also be applied to describe local spin environment in disordered atomic systems. For example, it could be used to develop a ML model for the dynamics of spin glasses, as exemplified by the dilute magnetic alloys such as CuMn. The interaction between the randomly distributed magnetic atoms in spin glass is mediated by conducting electrons of the metallic matrix. Conventionally, this effective spin-spin interaction is modeled by integrating out electrons beforehand, giving rise to the Ruderman-Kittel-Kasuya-Yosida (RKKY) pair interaction at weak coupling. The above magnetic ACSF descriptor combined with a learning model can provide large-scale dynamical simulations of spin-glass alloys with a better accuracy even for strong electron-strong coupling.  Another application is to combine the magnetic ACSF descriptor with molecular dynamics for the simulations of magnetic molecules in, e.g. ferrofluids~\cite{clark13,mertelj13}.

\begin{figure}
\includegraphics[width=1.0\columnwidth]{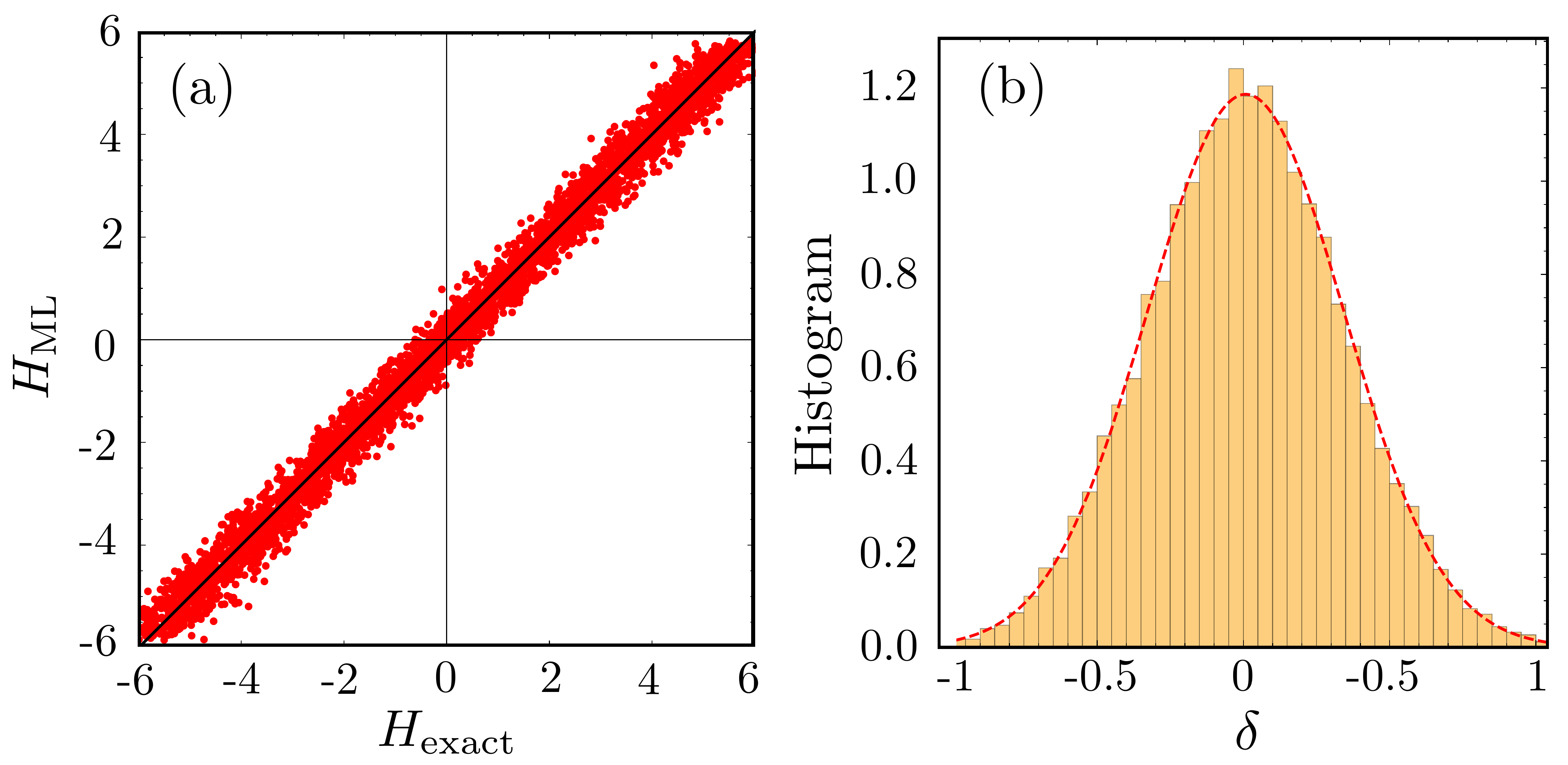}
\caption{(a) Exchange forces predicted by the ML model with the ACSF descriptor versus the exact solution from the exact diagonalization for the s-d model with exchange coupling $J = 6 t$. (b) Histogram of the force prediction error.}
\label{fig:DE-scsf}  
\end{figure}

Here we apply the descriptor to the case of s-d model on a square lattice. A two-parameter function of the form Eq.~(\ref{eq:F2-envelop})is used for both $F_2$ and $F_2'$. Specifically we use $F_2(R; d, w) = \exp[-(R - d)^2/w^2]$ to extract spin correlations at lattice sites in a ring of width $\sim \xi_2$ at a distance $\xi_1$ from the center site; a hard cutoff at $R_c$ is used.  On the other hand, since all triplet angles $\theta_{ijk}$ are pre-defined on a lattice, we ignore the angular dependence of the $F_3$ function. Instead of Eq.~(\ref{eq:F3-envelop}), we use the following parameterization:
\[
	F_3(R_{ij}, R_{ik}, R_{jk}) = e^{-\frac{(R_{ij} - d)^2 + (R_{ik} - d)^2}{w^2}} \,e^{-\frac{(R_{jk} - d')^2}{w'^2}}.
\]
In both 2- and 3-body symmetry functions, a finite width $w$ is used to ensure overlaps between consecutive rings, thus avoiding the spurious symmetry due to independence of rings.  Combining the ACSF descriptor with a NN model for the s-d model, Fig.~\ref{fig:DE-scsf} shows the ML predicted exchange forces versus the exact values. Although the ML prediction overall agrees well with the exact calculation, a rather large error was obtained with the ACSF descriptor, especially compared with the prediction error of the ML model with the bispectrum descriptor and the same NN structure to be discussed below. This is partly due to the fact that, from the representation theory point of view,  the ACSF descriptor is mostly dominated by the fully symmetric $A_1$ representation, as also discussed in Sec.~\ref{sec:scsf}. Contributions from other nontrivial IRs could be implicitly included in ACSF of higher order. For example, consider the 4-body symmetry functions (not included in our implementation)
\begin{eqnarray}
	G_{4} = \sum_{jkl \neq i} F_4(R_{ij}, \cdots) (\mathbf S_i \cdot \mathbf S_j)(\mathbf S_k \cdot \mathbf S_l).
\end{eqnarray}
This is a summation of the $A_1$ IR of four-spin variables.  As shown in \ref{table:D4-product}, this could result from the direct product of two-spin IR of either $A_2$, $B_1$, $B_2$, or $E$ symmetries, which cannot be captured by either of the two-spin symmetry functions in Eq.~(\ref{eq:G2b-spin}) and~(\ref{eq:G3a-spin}).

\subsection{Bispectrum}

\begin{figure}
\includegraphics[width=1.0\columnwidth]{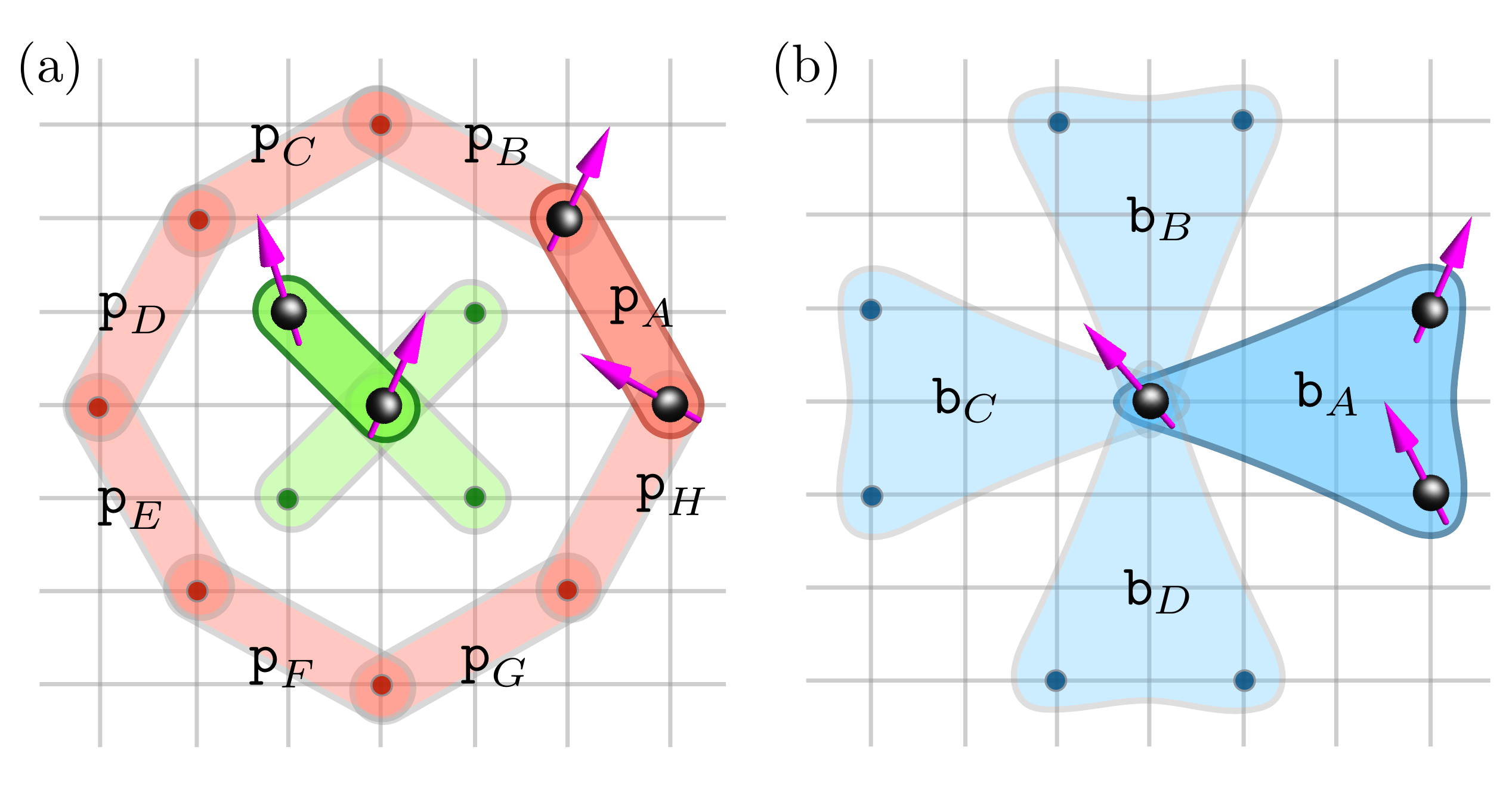}
\caption{Examples of (a) bond variables $\texttt{p}_{jk}$ and (b) spin scalar chirality $\texttt{b}_{ijk}$ that are related by the  point group symmetry of the lattice. These 2-spin and 3-spin correlations are also invariants of the internal SO(3) symmetry, and are building blocks for computing the bispectrum coefficients of the point group.}
\label{fig:spin-blk}  
\end{figure}

Compared with the ACSF, the bispectrum coefficients based on the group-theoretical method provides a more systematic approach to build the descriptor. Following the discussion in Sec.~\ref{sec:type2}, we consider a vector  $\vec{\mathcal{U}} = \{ \texttt{p}_{jk}, \texttt{b}_{jkl} \}$ consisting of the bond  and scalar chirality variables. which are already invariant with respect to the internal SO(3) rotations. This vector, which is a high-dimensional representation of the point group associated with a given center site, is then decomposed into the IRs. As discussed above, the collection of all such variables is an over-complete representation. A proper downsizing to avoid too much overlap is required, also for practical reasons. To this end, we restrict ourselves to three types of invariants shown in Fig.~\ref{fig:spin-blk}: (i) bond variables $\texttt{p}_{ij}$ between center site and another site-$j$ in the neighborhood, (ii) bond-variables $\texttt{p}_{jk}$ between two sites that are different from the center, and (iii) scalar chirality variables from triplets $(ijk)$ that include the center spin. Notably, they correspond to those used in the symmetry functions $G_{2b}$, $G_{3a}$, and $G_{3b}$, respectively, discussed in the previous subsection.

Even with this simplification, there are still a large number of these local variables. Fortunately, as discussed above, the vector representation for $\vec{\mathcal{U}}$ is block-diagonalized with each block consisting of spin-pairs or triplets with a fixed distance from the center site. For the case of square lattice, that means each block has a dimension of 4 or 8; there are also 12-member block, but these are trivial union of the dim-4 and dim-8 blocks. Examples of these blocks that are close on themselves under lattice rotation/reflection are shown in Fig.~\ref{fig:spin-blk}. The decomposition of these blocks is similar to that discussed in Sec.~\ref{sec:Q-bispectrum} for the scalar field. For the dimension-4 blocks, such as the four scalar chirality variables shown in Fig.~\ref{fig:spin-blk}(b), we have $4 = A_1 \oplus B_1 \oplus E$:
\begin{eqnarray} 
\label{eq:decomp-b}
f^{A_1} & = \texttt{b}_{A} + \texttt{b}_{B} + \texttt{b}_{C} + \texttt{b}_{D}, \nonumber \\
f^{B_1} & = \texttt{b}_{A} - \texttt{b}_{B} + \texttt{b}_{C} - \texttt{b}_{D}, \\
\bm f^E & = (\texttt{b}_{A} - \texttt{b}_{B}, \  \texttt{b}_{C} - \texttt{b}_D). \nonumber
\end{eqnarray} 
And the decomposition of a dim-8 block is illustrated by the eight off-center bond variables in Fig.~\ref{fig:spin-blk}(a): $8 = A_1 \oplus A_2 \oplus B_1 \oplus B_2 \oplus 2 E$. The corresponding expansion coefficients are
\begin{eqnarray} 
\label{eq:decomp-p}
 f^{A_1}  &=& \texttt{p}_{A} + \texttt{p}_{B} + \texttt{p}_{C} + \texttt{p}_{D} + \texttt{p}_{E} + \texttt{p}_{F} + \texttt{p}_{G} + \texttt{p}_{H}, \nonumber \\
 f^{A_2}  &=& \texttt{p}_{A} - \texttt{p}_{B} + \texttt{p}_{C} - \texttt{p}_{D} + \texttt{p}_{E} - \texttt{p}_{F} + \texttt{p}_{G} - \texttt{p}_{H}, \nonumber \\
 f^{B_1}  &=& \texttt{p}_{A} - \texttt{p}_{B} - \texttt{p}_{C} + \texttt{p}_{D} + \texttt{p}_{E} - \texttt{p}_{F} - \texttt{p}_{G} + \texttt{p}_{H}, \nonumber \\
 f^{B_2}  &=& \texttt{p}_{A} + \texttt{p}_{B} - \texttt{p}_{C} - \texttt{p}_{D} + \texttt{p}_{E} + \texttt{p}_{F} - \texttt{p}_{G} - \texttt{p}_{H}, \nonumber  \\
 \bm f^{E}  &=& (\texttt{p}_{A} - \texttt{p}_{E},\  \texttt{p}_{C} - \texttt{p}_{G}), \nonumber  \\
 \bm f^{E'}  &=& (\texttt{p}_{B} - \texttt{p}_{F},\  \texttt{p}_{D} - \texttt{p}_{H}).
\end{eqnarray}
Applying these decompositions to all blocks in the neighborhood, one obtains the coefficients of all IRs $\{ \bm f^{\Gamma } \}$  from the reducible representation of bond and scalar chirality variables $\vec{\mathcal{U}} = \{ \texttt{p}_{jk}, \texttt{b}_{jkl} \}$ within the cutoff radius. Here the basis functions of a given irrep are arranged into a vector $ \bm f^{ \Gamma } = (f^{\Gamma }_{ 1}, f^{ \Gamma }_{ 2}, \cdots, f^{ \Gamma }_{ n_\Gamma} )$, where $\Gamma$ labels the symmetry of the IR, and $r$ enumerates the multiple occurrence of $\Gamma$ in the decomposition of $\vec{\mathcal{U}}$.

Feature variables that are invariant under the point group symmetry are given by the bispectrum coefficients computed from these expansion coefficients. Details of the calculation of the bispectrum with the aid of the reference IRs can be found in Sec.~\ref{sec:Q-bispectrum} for the case of square lattice. Here we outline the procedure from a different perspective. First, the power spectrum $ p^{\Gamma} = | \bm f^{\Gamma }|^2$ is obviously invariant under discrete symmetry operations of the point group. However, power-spectrum descriptor contains spurious symmetries since it does not take into account the fact that different IR need to transform consistently, instead of independently, under symmetry operations. For example, the angle $\cos\theta_{12} = (\bm f^{({E}, 1)} \cdot \bm f^{(E, 2)} ) / | \bm f^{(E, 1)} | \,| \bm f^{(E, 2)} |$ between the vectors of two doublet IR characterizes the relative orientation of the two doublets, and is also an invariant of the point group. Consequently, the relative phases between different IRs should also be included in the descriptor in addition to the power spectrum~\cite{ma19}.

\begin{figure}
\includegraphics[width=1.0\columnwidth]{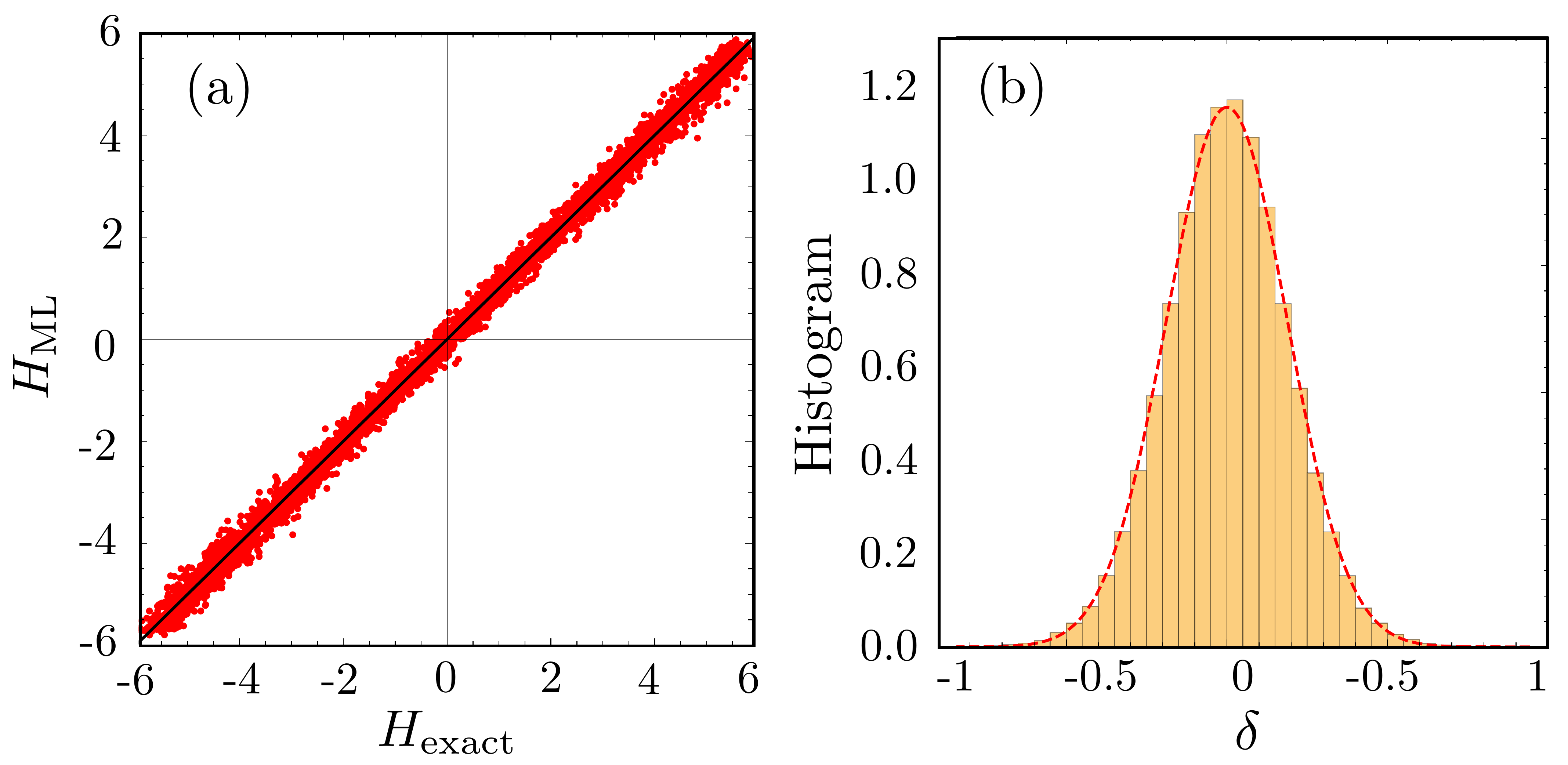}
\caption{(a) Exchange forces predicted by the ML model with the bispectrum descriptor versus the exact solution from the exact diagonalization for the s-d model with exchange coupling $J = 6 t$. (b) Histogram of the force prediction error.}
\label{fig:DE-bispectrum}  
\end{figure}

The reference IR $ \bm f^{(\mathbb{T}, *)}  $ discussed in Sec.~\ref{sec:ref-IR} offers a way to properly incorporate the ``phase" of the IRs into the descriptor.  These reference coefficients are computed by averaging large blocks of bond and chirality variables, such that they are less sensitive to small changes in the neighborhood spin configurations. We then define the relative ``phase" of a irrep as the projection of its basis functions onto the reference basis: $\eta^\Gamma  \equiv \bm f^\Gamma \cdot \bm f^\Gamma_{\rm ref} / |\bm f^\Gamma |\, |\bm f^\Gamma_{\rm ref}|$. The feature variables of the descriptor are then the collection of power spectrum coefficients and the relative phases: $ \{ p^{\Gamma} \,\, , \,\, \eta^\Gamma \}$. The various steps in the process of obtaining the descriptor are summarized in the following
\[
	\mathcal{C}_i \,\, \to \,\, \{ \texttt{p}_{jk}, \texttt{b}_{jkl} \} \,\, \to \,\, \{ \bm f^{\Gamma } \} \,\, \to \,\, \{ p^{\Gamma} \,\, ,  \eta^\Gamma \}
\]
The invariant feature variables characterizing the neighborhood spins, are then forwarded to the neural network which produces the local energy at its output node. This means the local energy associated with spin $\mathbf S_i$ depends on its neighborhood through the effective coordinates: $\varepsilon(\mathcal{C}_i) = \varepsilon(\{ p^\Gamma, \eta^\Gamma \}  )$, which obviously preserves both the SO(3) spin-rotational symmetry and the discrete lattice symmetry.

A six-layer NN model is constructed and trained using PyTorch~\cite{paszke19,nair10,barron17,paszke17,he15,kingma14}. The training dataset consists of 3500 snapshots of spins and local exchange forces, obtained from exact diagonalization of a $30\times 30$ lattice. Fig.~\ref{fig:DE-bispectrum}(a)~shows components of local exchange forces~$\mathbf H_i$ predicted by our trained NN model versus the exact results on test datasets. The difference $\delta = H_{{\text {ML}}} - H_{{\text {exact}}}$ is well described by a Gaussian distribution with a rather small mean-square error of $\sigma^2 = 0.035$, as shown in Fig.~\ref{fig:DE-bispectrum}(b). Interestingly, the histogram of the deviation $\delta$ implies that the statistical error of the ML model can be interpreted as an effective or artificial temperature in Langevin dynamics.

\section{Summary and discussion}
\label{sec:conclusion}

In this work, we present a numerical framework of utilizing machine learning methods for multi-scale dynamical modeling of condensed matter systems with emergent dynamical classical fields. These classical degrees of freedom could arise from the coupling to lattice dynamics, or magnetic moments of localized $d$ or $f$ electrons. They could also represent the collective electron behaviors, as exemplified by the order-parameter field in a symmetry breaking phase, of interacting electrons. The slow adiabatic dynamics of the emergent classical fields is often dominated by the electrons or quasi-particles, which are assumed to be in quasi-equilibrium of the instantaneous Hamiltonian parameterized by the classical variables. As in the quantum or {\em ab initio} molecular dynamics methods, accurate simulation of the dynamical classical fields requires solving the electronic structure problem at every time-step. Motivated by the success of ML-enabled large scale quantum MD simulations, we propose a similar approach for condensed matter systems in which the complex dependence of the local energy on the neighborhood classical field is encoded in a ML energy model. 

The two important components of the ML energy model are the descriptor for characterizing the local classical field configuration, and the learning model used to encoded the dependence on the local environment. Several learning models developed in the context of quantum MD can also be used for the effective energy model of the condensed-matter systems. Among the various ML models, the deep-learning neural network (NN) is perhaps the most versatile and accurate. The descriptor is crucial for properly incorporating symmetry of the system into the ML energy model. The so-called feature variables, which are input to the learning model, must be invariant with respect to symmetry transformations of the electron Hamiltonian. While a large number of descriptors have been proposed for ML-MD methods, the theory of descriptor for classical fields of condensed matter models has yet to be developed. 

We discuss common features of the descriptor of classical fields of electronic lattice models, and formulate a general theory by first distinguishing two types of models depending on the absence or presence of an internal symmetry for the classical fields.  Several specific approaches to derive a descriptor have been discussed. First, a general descriptor is given by the ordered eigenvalues of the correlation matrix of the neighborhood classical fields, which is similar in spirit to the Weyl or Coulomb matrix descriptor used to characterize the atomic environment. Another approach, also motivated by ML-models for quantum MD simulation, is the generalization of the atom-centered symmetry functions which incorporates the internal symmetry of the classical fields.

The majority of our effort focuses on the group theoretical method which offers systematic and controlled approach to build fundamental invariants of the symmetry group. In this approach, the local classical fields, which form a high-dimensional representation of the site-symmetry point group, is first decomposed into the irreducible representations. Fundamental invariants are given by the bispectrum coefficients of three IRs, which are similar to the scalar or triple product of three vectors. To cope with the issue due to the large number of the over-complete bispectrum invariants, we propose a simplification method based on the concept of the reference IRs. Instead of keeping all the bispectrum coefficients, both the amplitude and the relative ``phase" of each IR can be faithfully retained via an inner product with the reference IR. 

Finally, we demonstrate the implementation of the various descriptors on well-known lattice models including the Holstein and Jahn-Teller model, and the s-d Hamiltonian for itinerant magnets. The classical field in the former case corresponds to local structural distortions. In particular, the scalar field in the Holstein-type models offers the simplest example to illustrate the working of the lattice descriptor. On the other hand, the s-d model characterized by a vector classical field is used to demonstrate the construction of a descriptor with an independent internal symmetry.

Our work laid the foundation for applying ML methods to multi-scale dynamical modeling in condensed matter systems. Contrary to ML-based MD methods which is an ongoing active research field by itself, the goal here is to model the adiabatic dynamics of classical fields under the influence of quasi-equilibrium electrons. The capability of going beyond empirical methods for large-scale dynamical simulations of such classical fields has numerous implications in condensed matter physics. For example, one particularly important application is the accurate dynamical modeling of topological defects of multi-component classical fields, which are prevalent in condensed matter systems. Notable examples include vortices in superconductivity and skyrmions in itinerant magnetism. 

Moreover, complex inhomogeneous electronic states are ubiquitous in correlated electron systems. Not only are these mesoscopic textures of fundamental importance in correlated electron physics, they also play a crucial role in the emergence of novel macroscopic functionalities. For example, complex mixed-phase states are prevalent in colossal magnetoresistant materials and several high-$T_c$ superconductors also exhibit intriguing stripe or checkerboard patterns. Accurate modeling of these complex nanoscale textures is thus of paramount importance in the engineering of these novel material functionalities. However, large-scale simulations of such electronic textures so far are based on empirical or phenomenological models, mostly because of the extreme difficulty for the multi-scale dynamical modeling of such systems.   We believe that the ML force field approach along with the proper descriptor outlined in this work will be an indispensable tool to enable large-scale dynamical simulations of complex patterns in correlated electron materials.

\begin{acknowledgements}
This work was supported by the US Department of Energy Basic Energy Sciences under Award No. DE-SC0020330. The authors also acknowledge the support of Research Computing at the University of Virginia.
\end{acknowledgements}

\end{document}